\documentstyle[12pt]{article}

\newcommand{\newsection}{
\setcounter{equation}{0}
\section}
\newcommand{\tr}{\,{\rm tr}\,}
\def\e{{\,\rm e}\,}

\def\eop{\vspace*{\fill}\pagebreak}
\def\be{\begin{equation}}
\def\ee{\end{equation}}
\def\bea{\begin{eqnarray}}
\def\eea{\end{eqnarray}}

\def\Mdclose#1{{\overline {\cal M}}^{disc}_{#1}}
\def\aa{\alpha}
\def\bb{\beta}

\def\L{\Lambda}

\def\l{\lambda}
\def\h{\eta}

\def\t{\widetilde}

\def\g {\gamma}

\def\pip{\pi^{-1}(p_{\ast})}
\def\tpip{\widetilde\pi^{-1}(p_{\ast})}
\def\ep {\varepsilon}

\def\Mcomb{{\cal M}_{g,n}^{comb}}
\def\Mdisc{{\cal M}_{g,n}^{disc}}
\def\Mcdisc{{\overline {\cal M}}_{g,n}^{disc}}
\def\MgnR{{\cal M}_{g,n}\otimes {\bf R}_{+}^{n}}
\def\Mgn{{\cal M}_{g,n}}
\def\Mpar#1{{\cal M}_{#1}}
\def\Tgn{{\cal T}_{g,n}}
\def\Td{{\bf T}^d}
\def\Mc{{\overline {\cal M}}_{g,n}}
\def\Mcpar#1{{\overline {\cal M}}_{#1}}

\newcommand{\dd}[1]{{\partial \over \partial #1}}

\newcommand{\ra}{\rightarrow}

\newcommand{\LL}{{\cal L}}
\newcommand{\TT}{{\cal T}}

\newcommand{\OO}{{\cal O}}
\newcommand{\ZZ}{{\cal Z}}
\newcommand{\Lan}{\langle \!\langle}
\newcommand{\Ran}{\rangle \!\rangle}
\newcommand{\<}{\langle}
\renewcommand{\>}{\rangle}
\newcommand{\fr}[2]{{\textstyle {#1 \over #2}}}
\newcommand{\redefine}{\renewcommand}

\title{{\bf \mbox{} \\Matrix Models: a Way to Quantum Moduli Spaces}
\vspace{.5cm}}
\author{{\bf L. Chekhov}\thanks{E--mail: \  chekhov@lpthe.jussieu.fr
\ and \
chekhov@qft.mian.su}
\date{ }
\vspace{.5cm} \\
{\it Laboratoire de Physique Theorique et Hautes Energies} \\
{\it Universit\'e Paris-VI, Tour 16, 4 place Jussieu}\\
{\it Paris, France}}

\begin{document}

\maketitle

\vspace{-10.6cm}

\begin{flushright}
PAR--LPTHE--93--22 \\ April 21, 1993
\end{flushright}

\vspace{6.8cm}

\begin{abstract}
We give the description of discretized moduli spaces (d.m.s.) $\Mcdisc$
introduced in \cite{Ch1} in terms of a discrete de Rham cohomologies
for each
moduli space $\Mgn$ of a genus $g$, $n$ being the number of punctures. We
demonstrate
that intersection indices (cohomological classes) calculated for d.m.s.
coincide with the ones for the continuum moduli space $\Mc$ compactified
by
Deligne and Mumford procedure. To show it we use a matrix model technique.
The
Kontsevich matrix model is a generating function for these indices in the
continuum
case, and the matrix model with the potential
$N\alpha \tr {\bigl(- \fr 12 \L X\L X +\log (1-X)+X\bigr)}$
is the one for d.m.s.
In the last case the effects of reductions become relevant, but we use the
stratification procedure in order to express integrals over open spaces
$\Mdisc$ in terms of intersection indices which are to be calculated on
compactified
spaces.  The
coincidence of the cohomological classes for both continuum and
d.m.s. models enables us to
propose the existence of a quantum group structure on d.m.s. Then d.m.s.
are
nothing but cyclic (exceptional) representations of a quantum group
related
to a moduli space $\Mc$. Considering the explicit expressions for
integrals
of Chern classes over $\Mc$ and $\Mcdisc$ we conjecture that each moduli
space
$\Mc$ in the Kontsevich parametrization can be presented as a coset $\Mc
={\bf T}^d/G$, $d=3g-3+n$, where ${\bf T}^d$ is some $d$--dimensional
complex torus and $G$ is a finite order symmetry group of ${\bf T}^d$.

\end{abstract}

\eop

\newsection{Introduction.}

In our recent papers \cite{Ch1}, \cite{ACKM} we have proposed and developed an
approach to discretization of arbitrary moduli space of algebaric curve.
The
connection of these spaces to a matrix model was established and also it
was
demonstrated explicitly that in the limit when discretization parameter
became small this matrix model goes to the Kontsevich one \cite{Kon91}
This model, in its
turn,
is a generating function for intersection indices or
integrals of first Chern classes on the corresponding moduli space.
It was a proposal by Witten \cite{Wit1} that these integrals yield correlation
functions for the two--dimensional gravity coupled to the matter.
Still
the notion of the discretization of the moduli space was not very certain.
In
the present paper we ensure this object with some explicit description and,
using the results of \cite{ACKM}, we prove the main identity:
\be
\Lan \tau_{d_1}\dots\tau_{d_n}\Ran_g = \< \tau_{d_1}\dots\tau_{d_n}\>_g,
\label{i1}
\ee
where in both cases $\<\tau_{d_1}\dots\tau_{d_n}\>_g$ are someway determined
integrals of products of Chern classes $\omega_i$  and $\t\omega_i$ over
moduli spaces. (We shall denote as tilde--variables the discrete counterparts
of the
continuum objects.) In the continuum case
\be
\<\tau_{d_1}\dots\tau_{d_n}\>_g=\int_{\Mc}\prod_{i=1}^n\omega_i^{d_i}.
\label{i2}
\ee
On the l.h.s. of (\ref{i1}) $\Lan
\tau_{d_1}\dots\tau_{d_n}\Ran_g $ can be presented in a form similar to
(\ref{i2}) but with all quantities being related to the discretized moduli
space (d.m.s.). In what follows we shall provide a proof to (\ref{i1})
using
a matrix model technique.

We shall use a fat graph technique in order to introduce coordinates on the
moduli
spaces. The coordinatization means that we assign lenths $l_i$ to all
edges
of the fat graph and the number of punctures, $n$, is  the
number
of faces of the graph. We call this space $\Mcomb$. The discretization of
$\Mcomb$ is rather simple -- we assume that all these lengths are to be
integer numbers (probably zeros). When taking all these possibilities
we get the points of the discretized moduli space $\Mcdisc$.
For a general oriented graph of the genus $g$ and the number of faces $n$
the total number of edges (for trivalent vertices of the general position)
is $6g-6+3n$ which exceeds the dimension of $\Mgn$ by $n$. So there are $n$
extra parameters which are not related to  the coordinates on the
original moduli
space itself. Namely, they are perimiters of the faces of the graph. In the
continuum case we have due to Strebel theorem \cite{Streb} an isomorphism
$\MgnR \simeq \Mcomb$ and we define a projection $\pi : \Mcomb\to {\bf R}_+^n$
to
the space of perimeters. The fibers $\pi^{-1}(p_{\ast})$ of the inverse
map
are isomorphic to the initial moduli space $\Mgn$ and hence they all are
isomorphic to each other. We are able now also to define another
projection
$\t\pi : \Mdisc\to {\bf Z}_+^n|_{\sum p_i\in 2{\bf Z}_+}$ where all
perimeters
are strictly positive
integers with even total sum and consider its fibers
$\t\pi^{-1}(p_{\ast})$. They are, generally speaking, finite sets of points
belonging to the initial moduli space $\Mc$. These sets are no more
isomorphic to each other. Moreover, among these points there are always
points which correspond exactly to reduced surfaces (``infinity
points'').
We assume that the space of reduced surfaces is $\partial\Mgn = \Mc-\Mgn$.
In the usual Teuchm\"uller picture all these points lie at the infinity, but in
what
follows we should include them into the game.
We
are able to introduce an analogue of De Rham complex on these spaces using
finite difference structures instead of differential ones. There are the
spaces we call discretized moduli spaces (d.m.s.). Also instead of
$U(1)$--bundles for continuum case we shall consider ``${\bf
Z}_p$--bundles'' over these spaces. Thus we can define cohomological
classes for d.m.s. as well
\be
\Lan \tau_{d_1}\dots\tau_{d_n}\Ran_g =
\int_{\tpip}\prod_{i=1}^n\t\omega^{d_i}_i.
\label{i2a}
\ee
There is an unique (up to isomorphisms) closed moduli space $\Mc=\pip
[\Mcomb]$ and an infinite series of nonisomorphic $\tpip [\Mcdisc]$, but
for
all of them the relation (\ref{i1}) holds true. We shall present matrix
model arguments in favour of this statement, but right now we want to
discuss consequences of the relation (\ref{i1}).

{}From the one hand, it means that there exists an invariant which doesn't
depend on which space from the set $\{\pip,\tpip\}$ we choose. Since we
replace the differential structure for $\pip$ by a discrete one for
$\tpip$,
it is natural to look for an analogy with the Quantum
Group (QG) description for quantum deformations. Then for any $\tpip$
we expect that the
space of functions $f[\tpip]$ is a representation of the
Quantum Group $Q_{g,n}$ related to the initial moduli space $\Mc$. This
representation has a feature that it does not contain highest weight vectors,
hence it should be a cyclic (exceptional) representation of QG
\cite{Sklyan}, \cite{Kac}. These representations exist only for specific values
of
quantum deformation parameters $q_j=\e^{2\pi i m_j/s_j}$, $(m_j,s_j\in
{\bf
Z})$,  the values of $n_j$ being related to the set of perimeters
$\{p_{\ast}\}$. In particular, for $\Mc$ with $n=1$ (one puncture)
$s_j=p_1/2$. Then the intersection indices themselves are values of
some invariant traces of quantum operators acting on these representations.
It is worth to note that due to orbifold structure of the moduli
space $\Mc$ we can view it as a coset of some finite covering $\Tgn$ of
$\Mc$ over a symmetry group $G$: $\Mc=\Tgn/G$. It means that these
representations themselves possess this internal symmetry group.

{}From the other hand, we shall show that in the Kontsevich's
parametrization
the very evaluation of
the integrals over $\pip$ and $\tpip$ can be reduced to a calculation
of integrals over volume forms on abovementioned finite coverings $\Tgn$.
The discretization on this language means that we introduce an equidistant
lattice on $\Tgn$ and while calculating the volume we merely count a total
number of sites in this lattice and divide it by some product of $p_i^2$:
$p_1^{2a_1}\dots p_n^{a_n}$ where $\sum_{i=1}^na_i=d=3g-3+n$ -- the total
dimension of $\Mc$.
Note also that all $\Tgn$ are compact spaces without
boundaries. One may imagine that when doing the sum over all points of
the
lattice they contribute to this sum together with unit cubes therefore
calculating the volume of $\Tgn$. But it is true only if all these points
are nonsingular points, i.e. of zero curvature! Because of a huge variety of
sets
$\tpip$ we know that for each orbifold point there exists a set
$\{p_1,\dots,p_n\}$ which includes this point. Were there exists a
singularity of metric in this point,
then the equality (\ref{i1}) is broken, but it contradicts
to our proof of its validity
in which we use
only matrix model arguments without any reference to an underlying
geometrical structure. It means that there is no points of nonzero
internal
curvature in the space $\Tgn$, hence $\Tgn$ should be a torus of the
complex dimension $d$, and in the Kontsevich parametrization:
\be
\Mc = \Td /G,
\label{i2b}
\ee
where $\Td$ is a $d$--dimensional complex torus and $G$ is a finite order
symmetry group of $\Td$. In particular, the order of this group is a
common
denominator for all intersection indices on this moduli space.

In order to find a connection between moduli spaces $\Mc$ and d.m.s. we
shall use a matrix model technique. Matrix models recently revealed
a lot of applications in various
branches
of mathematical physics: two--dimensional quantum field theory,
intersection
theory on the moduli space of Riemann surfaces, etc.

An old days concept for usual 1--matrix hermitian model was the following:
in a ``fat graph'' technique starting with each graph we can
construct the dual one corresponding to some Riemann surface
with singularities of curvature concentrated in vertices of this dual
graph. Then faces of this graph correspond to vertices in the matrix model
graph and vice versa. If the initial potential
contains only three valent vertices we
can speak about ``triangulation'' of the Riemann surface. In what follows
we shall deal with potentials of an arbitrary order, but we use the same
term
``triangulation''. The model with an arbitrary potential was solved exactly in
\cite{BK90} in the double scaling limit when
the number
of triangles tends to infinity and these singular
metrics approximate ``random
metrics'' on the surface. These model was  presented by a
hermitian $N\times N$ one--matrix model
\be
\int \exp\bigl( \tr P(X) \bigr) DX,
\label{herm}
\ee
where $P(X)=\sum _{n}T_n \tr X^n$, $T_n$ being times for the
one--matrix model.
For such system discrete Toda chain equations holds with an additional
Virasoro symmetry imposed \cite{KawMMM}. In the limit $N\to\infty$ the
Korteveg--de--Vries equation arises. The partition function of the
two--dimensional gravity for this approach is a series in an infinite
number
of variables and coincides with the logarithm of some $\tau$--function for
KdV hierarchy.

Another approach to the two--dimensional gravity is to do the integral over
all classes of conformally equivalent metrics on Riemann surfaces. It may
be
presented as an integral over the finite--dimensional space of conformal
structures. This integral has a cohomological description as an
intersection
theory on the compactified moduli space of complex curves. Edward Witten
presented compelling evidence for a relationship between random surfaces
and
the algebraic topology of moduli space \cite{Wit1}, \cite{Wit2}. In fact,
he
suggested that these expressions coincide since both satisfy the same
equations of KdV hierarchy. It was Maxim Kontsevich who proved this
assumption \cite{Kon91}. Surprisingly, he explicitly presented a new matrix
model defining exactly the values of intersection indices or, on the
language
of 2D gravity, correlation functions of observables $\OO _n$ of the type
\be
<\OO _{n_1}\dots \OO_{n_s}>_g,
\ee
where $<\dots >_g$ denotes the expectation value on a Riemann surface with
$g$ handles. Then the string partition function $\tau (t)$ has an
asymptotic
expansion of the form
\be
\tau (t)=\exp \sum_{g=0}^{\infty}\left\langle \exp \sum_{n}t_n \OO _n
\right\rangle_{g},
\label{taufunct}
\ee
and it is certainly a tau--function of the KdV hierarchy taken at a point of
Grassmannian where it is invariant under the action of the set of the
Virasoro constraints: ${\cal L}_n \tau (t)=0$, $n\ge -1$ \cite{Fuk},
\cite{Wit3}, \cite{GN}, \cite{MMM}. One might say that the Kontsevich model
is used to triangulate moduli space, whereas the original models
triangulated
Riemann surfaces (see e.g. \cite{Dij91}).

The generalization of the Kontsevich model --- so-called Generalized
Kontsevich Model (GKM) \cite{KMMMZ} is related to the two--dimensional Toda
lattice hierarchy and it originated from the external field problem defined
by the integral
\be
Z[\L ;N]=\int DX \exp \left\{N \tr {\bigl( \L X -V_0 (X)\bigr)}\right\},
\label{external}
\ee
where $V_0(X)=\sum_{n} t_n \tr X^n$ is some potential, $t_n$ are related to
times of the hierarchy. This model is equivalent to the Kontsevich integral
for $V_0(X) \sim \tr X^3$. To solve the integral (\ref{external}) one may
use
the Schwinger--Dyson equation technique \cite{BN81} written in terms of
eigenvalues of $\L $. The Kontsevich model was solved in the genus
expansion
in the papers \cite{GN}, \cite{MakS} for genus zero (planar diagrams) and
in
\cite{IZ92} for higher genera.

Recently, the Kontsevich--Penner model was introduced \cite{CM92a}.
The Lagrangian of this  model has the following form:
\be
{\cal Z}[\Lambda] = \int DX \exp \left( N \,\hbox{tr} \left\{
-\frac 12\Lambda X\Lambda X
+\alpha \bigl[ \log(1+X)-X\bigr]\right\}\right),\,\ \ \Lambda =
\hbox{ diag} (\Lambda_1, \dots, \Lambda_N).
\label{PK}
\ee
This model may
be readily reduced to (\ref{external}) with $V_0(X)= -X^2/2+\alpha \log X$.
It was solved in genus expansion in \cite{CM92a}, \cite{ACM}. It appears
(see \cite{CM92b}, \cite{KMMM}) that it is in fact equivalent to the
one--matrix
hermitean model (\ref{herm}) with the general potential
\be
P(X)=\sum_{n=0}^{\infty}T_n \tr X^n,
\ee
which times are defined by the kind of Miwa transform
$(\eta=\L-\alpha\L^{-1})$:
\be
T_n=\frac 1n \tr\eta ^{-n} -\frac N2 \delta _{n2}\ \hbox{ for }\ n\ge 1
\ \ \hbox{ and }\ T_0=\tr \log \eta ^{-1}.
\ee
It was demonstrate in \cite{Ch1}, that this new model describes in a
natural
way the intersection indices for the case of d.m.s. The only
complification
is that this  model does not present generating function for the
indices (\ref{i2a}) straightforwardly because of contribution from
reductions. Indeed, any matrix model can deal with only open
strata of a moduli
space. It was not essential for the case of the Kontsevich model since
there the integration went over cells of the highest dimension in the
simplicial complex partition of the moduli space $\Mc$. All singular
points
are simplices of lower dimensions in $\Mc$ and give no
contribution
to the integral. But in the case of d.m.s. integrals over simplices of all
dimensions are relevant due to the total discretization, so the
integrals
over reduced surfaces give
non--zero contribution which we should exclude in order to compare with
the
matrix model. The way to do it is to use a stratification procedure
\cite{Mum83} which permits to express open moduli space $\Mgn$ via $\Mc$ and
moduli spaces of lower genera.

All this reveals one new side of seems--to--be--well--known standard
one--matrix model.
This model appears to have the direct relation to $2D$--gravity not only in the
d.s.l.
but also beyond it. On this language d.s.l. means that we introduce using the
Laplace
transform a small parameter $\ep$ and while $\ep$ goes to zero the leading
contribution to the sum over d.m.s. originates from d.m.s. with large values of
$p_i$,
i.e. from d.m.s. with more and more dense distribution of points of d.m.s.
inside the
original moduli space $\Mcdisc$. Note that on the language of quantum
description $\ep$
plays the role of the Plank constant $\h$.

In the paper \cite{ACKM} the explicit solution to the Kontsevich--Penner,
or, equivalently, to the general one--matrix model was found in genus
expansion. The key role in this consideration was played by so-called
``momenta'' of the potential resembling in many details ``momenta'' which
appeared in genus expansion solution to the Kontsevich model \cite{IZ92}. We
shall use some proper reexpansion of these momenta in terms of
new quantities which stand just by the intersection indices (\ref{i2}),
(\ref{i2a}). These new variables are quantum analogues of the Kontsevich
``times''
$T_n=\tr \L^{-2n-1}$. Summing up we stress that the usual one--matrix model
appears to be still rather interesting even beyond the double
scaling
limit, and it may provide a way to quantization of
the
moduli spaces.

The paper is organized as follows: The short review of geometric approach to
the
Kontsevich model is given in Section~2. The Penner--Kontsevich model as well as
the
definition of the discretized moduli spaces are presented in Section~3. In
Section~5
the matrix model technique is actievely used in order to compare both models
and it is
shown by this comparison that the intersection indices in both models coincide.
The
proposal for the orbifold structure of modular spaces in the Kontsevich
parametrization
is the matter of Section~5. Also we discuss there the possibility to introduce
Quantum
Group structures on these discrete moduli spaces. Then the short Conclusion
section
is followed by Appendix in which we present an explicit solution for the
modular space
$\Mcpar{2,1}$ of genus 2.

\newsection{The geometric approach to the Kontsevich model.}

In his original paper \cite{Kon91} Kontsevich proved that
\be
\sum_{d_1,\dots,d_n=0}^{\infty}<\!\tau_{d_1},\tau_{d_2},\dots,\tau_{d_n}\!>
\prod_{i=1}^{n}(2d_i-1)!!\l _i^{-(2d_i+1)}=
\sum_{\Gamma}{2^{-\# X_0}\over \#\hbox{\,Aut\,}(\Gamma )}
\prod_{\{ij\}}{2\over \l_i+\l_j},
\label{Aut}
\ee
where the objects standing in angular brackets on the left--hand side are
(rational) numbers describing intersection indices, and on the right--hand
side the sum runs over all oriented connected trivalent ``fatgraphs''
$\Gamma$ with $n$ labeled boundary components, regardless of the genus,
$\# X_0$ is the number of vertices of $\Gamma$, the product runs over all the
edges in the graph and $\#\hbox{\,Aut}$ is the volume of discrete symmetry
group of the graph $\Gamma$.

The amazing result by Kontsevich is that the quantity on the right hand
side of (\ref{Aut}) is equal to a free energy in the following matrix
model:
\be
\e^{F_N(\L)}={\int dX\,\exp\left(-\frac 12 \tr \L X^2+\frac 16 \tr
X^3\right)
\over \int dX\,\exp\left(-\frac 12 \tr \L X^2 \right)},
\label{Konts}
\ee
where $X$ is an $N\times N$ hermitian matrix and $\L=\hbox{\,diag\,}(\l_1,
\dots ,\l_N)$. The distinct feature of the expression (\ref{Aut}) is that
in spite of the fact that each selected diagram has quantities
$(\l_i+\l_j)$ in the
denominator, when taking a sum over all diagrams of the same genus and
the same number of boundary components all these quantities are cancelled
with the ones from nominator.

Feynman rules for the Kontsevich matrix model are the following: as in the
usual matrix models, we deal with so-called ``fat graphs'' or ``ribbon
graphs'' with propagators having two sides, each carries corresponding
index.
The Kontsevich model varies from the standard one--matrix hermitian model
since there appear additional variables $\l_i$ associated with index
loops in the diagram, the propagator being equal to $2/(\l_i+\l_j)$, where
$\l_i$ and $\l_j$ are variables of two cycles (perhaps the same cycle)
which the two sides of propagator belong to. Also there are trivalent
vertices presenting the cell decomposition of the moduli space.

It is instructive to consider the simplest example
of genus zero and three boundary components which we symbolically label
$\l_1$, $\l_2$ and $\l_3$. There are two kinds of diagrams giving the
contribution in this order (Fig.1).
The contribution to the free energy arising from this sum is
\bea
&{}& {1\over 6(\l_1+\l_2)(\l_1+\l_3)(\l_2+\l_3)}+\frac 13 \biggl\{
{1\over 4\l_1(\l_2+\l_1)(\l_3+\l_1)}\biggr. + \nonumber\\
&+&\biggl.(1\ra 2,2\ra 3, 3\ra 1)+
(1\ra 3,3\ra 2,2\ra 1)\biggr\}\nonumber\\
&=&{2\l_1\l_2\l_3+\l_2\l_3(\l_2+\l_3) +\l_1\l_3(\l_1+\l_3)
+\l_1\l_2(\l_1+\l_2)\over 12\l_1\l_2\l_3(\l_1+\l_2)(\l_1+\l_3)(\l_2+\l_3)}
\nonumber\\
&=&{1\over 12\l_1\l_2\l_3}.
\eea
This example demonstrates the cancellations of $(\l_i+\l_j)$--terms in
the denominator above mentioned.

\eop

\phantom{AAAAAAAAAAAAA}

%
%
\begin{picture}(190,2)(-50,85)

\put(40,40){\oval(40,40)[b]}
\put(40,40){\oval(30,30)[b]}
\put(40,45){\oval(40,40)[t]}
\put(40,45){\oval(30,30)[t]}
\put(20,40){\line(0,1){5}}
\put(60,40){\line(0,1){5}}
\put(25,40){\line(1,0){30}}
\put(25,45){\line(1,0){30}}
\put(3,40){\makebox{1/6}}
\put(20,65){\makebox{$\lambda_1$}}
\put(35,50){\makebox{$\lambda_2$}}
\put(35,30){\makebox{$\lambda_3$}}
\put(63,40){\makebox{+1/2}}
\put(100,40){\oval(30,30)[b]}
\put(100,40){\oval(20,20)[b]}
\put(100,45){\oval(30,30)[t]}
\put(100,45){\oval(20,20)[t]}
\put(85,40){\line(0,1){5}}
\put(90,40){\line(0,1){5}}
\put(110,40){\line(0,1){5}}
\put(115,40){\line(1,0){20}}
\put(115,45){\line(1,0){20}}
\put(150,40){\oval(30,30)[b]}
\put(150,40){\oval(20,20)[b]}
\put(150,45){\oval(30,30)[t]}
\put(150,45){\oval(20,20)[t]}
\put(165,40){\line(0,1){5}}
\put(160,40){\line(0,1){5}}
\put(140,40){\line(0,1){5}}
\put(168,40){\makebox{+perm.}}
\put(125,55){\makebox{$\lambda_1$}}
\put(95,40){\makebox{$\lambda_2$}}
\put(145,40){\makebox{$\lambda_3$}}
\end{picture}

\vspace{5cm}

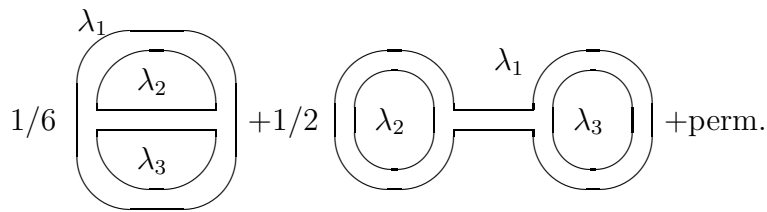
\centerline{Figure 1.  the g=0, s=3 contribution to Kontsevich's model}

\vspace{6pt}

\subsection{Geometry of fiber bundles on $\Mc$.}

Now the sketch of Kontsevich's proof is in order. Let us  associate  with
each edge $e_i$ of a fat  graph  its  length  $l_i>0$.  We  consider  the
orbispace $\Mcomb$ of fat graphs with all possible lengths of edges and
arbitrary valencies of vertices. Two graphs are equivalent if an
isomorphism between them exists. Let us introduce an important object ---
the space of $(2,0)$--meromorphic differentials $\omega (z)dz^2$ on a
Riemann surface with $g$ handles and $n$ punctures, the only poles of
$\omega (z)$ are $n$ double poles placed in the points of punctures
with strictly positive quadratic residues $p_i^2>0$, $(i=1,\dots,n)$. It
is Strebel's theorem \cite{Streb} which claims that the natural mapping
from $\Mcomb$ to the moduli space $\MgnR$, where ${\bf R}_+^n$ is the space
of residues, $p_i>0$ being perimeters of cycles, is
homeomorphism. Thus, varying $l_j$ and taking the composition of all
graphs we span the whole space $\MgnR$.

Each cycle can be interpreted as a boundary components $I_i$ of the
Riemann surface since in the Strebel metric it can be presented as
half-infinite cylinder with the puncture point placed at infinity. The
boundary of it consists of a finite number of intervals (edges). We
consider a set of line bundles ${\cal L}_i$ which fiber at a point
$\Sigma\in\Mgn$ is the cotangent space to the puncture point
$x_i$ on the surface $\Sigma$. The
first Chern class $c_1({\cal L}_i)$ of the line bundle ${\cal L}_i$
admits a representation
in
terms of the lengths of the intervals $l_j$.
The perimeter of the boundary component is $p_i=\sum_{l_\alpha\in
I_i}l_\alpha$.

The first step in constructing $c_1({\cal L}_i)$  is  to  determine
$\alpha_i$ which  is  $U(1)$--connection on the boundary component
corresponding to $i$th
puncture. In order to explicitly describe this construction it is
convenient
to introduce ``polygon  bundles'' $BU(1)^{comb}_{(i)}$   in   Kontsevich's
notations. These polygon bundles are sets of equivalent classes of all
sequences of positive real numbers $l_1,\dots , l_k$ modulo cyclic
permutations.

$BU(1)^{comb}$ is the moduli (orbi)space of numbered ribbon graphs with
metric whose underlying graphs are homeomorphic to the circle. There is an
$S^1$--bundle over this orbispace whose total space $EU(1)^{comb}$ is an
ordinary space. The fiber of the bundle over the equivalence class of
sequences $l_1,\dots, l_k$ is a union of intervals of lengths
$l_1\dots, l_k$ with pairwise glued ends, i.e. a polygon.
One may prove that the map $\MgnR \to \bigl( BU(1)^{comb}\bigr)^n$ extend
continuously to $\Mc\times{\bf R}_+^n$. The inverse images of
$S^1$--bundles
are naturally isomorphic to the circle bundles associated with the complex
line bundles $\LL_i$.

Let us now compute the first Chern class of the circle bundle on
$BU(1)^{comb}$. The points of $EU(1)^{comb}$ can be identified with pairs
$(p,S)$ where $p$ is a perimeter and $S$ is a nonempty finite subset
(vertices) of the circle ${\bf R}/p{\bf Z}$. Denote $0\leq \phi_1<\dots
<\phi_k <p$ representatives of points of $S$. The lengths of the edges of
the
polygon are
\be
l_i=\phi_{i+1}-\phi_i\quad (i=1,\dots,k-1),\quad l_k=p+\phi_1-\phi_k.
\label{k1}
\ee
Now we should choose some convenient form for $S^1$--connections on these
polygon bundles.
Denote by $\alpha$ the 1-form on $EU(1)^{comb}$ which is equal to
\be
\alpha = \sum_{i=1}^k \frac{l_i}{p}\times d\left(\frac{\phi_i}{p}\right).
\label{k2}
\ee
Certainly $\alpha$ is well--defined and the integral of it over each fiber
of
the universal bundle $EU(1)^{comb}\to BU(1)^{comb}$ is equal $-1$. The
differential $d\alpha$ is the pullback of a 2--form $\omega$ on the base
$BU(1)^{comb}$,
\be
\omega = \sum_{1\leq i< j\leq k-1} d\left(\frac{l_i}{p}\right)\wedge
d\left(\frac{l_j}{p}\right).
\label{k3}
\ee
Extrapolating these results to the compactified moduli spaces we obtain
that
the pullback $\omega_i$ of the form $\omega$ under the $i$th map
$\Mc\times{\bf R}_+^n\to BU(1)^{comb}$ rerpesents the class $c_1(\LL_i)$.

Denote by $\pi: \Mcomb\to {\bf R}_+^n$ the projection to the space of
perimeters. Intersection indices are given by  the formula:
\be
\<\tau_{d_1}\dots\tau_{d_n}\>=\int_{\pi^{-1}(p_{\ast})}\prod_{i=1}^n
\omega_i^{d_i},
\label{k4}
\ee
where $p_{\ast}=(p_1,\dots, p_n)$ is an arbitrary sequence of positive real
numbers and the $\pi^{-1}(p_{\ast})$ is a fiber of $\Mc$ in $\Mcomb$.

\subsection{Matrix Integral}
We denote by $\Omega$ the two--form on open strata of $\Mcomb$:
\be
\Omega = \sum_{i=1}^{n} p_i^2 c_1({\cal L}_i),
\label{omega}
\ee
which restriction to the fibers of $\pi$ has constant coefficients in the
coordinates $\bigl(l(e)\bigr)$. Denote by $d$ the complex dimension of
$\Mgn$, $d=3g-3+n$. The volume of the fiber of $\pi$ with respect to
$\Omega$
is
\bea
\hbox{vol}\bigl(\pi^{-1}(p_1\dots ,p_n)\bigr) &=&
\int_{\pi^{-1}(p_{\ast})} {\Omega ^d\over d!} = \frac{1}{d!}
\int_{\pi^{-1}(p_{\ast})}\bigl(p_1^2c_1({\cal L}_1)+\dots +
p_n^2c_1({\cal L}_n)\bigr)^{d}=\nonumber\\
&=&\sum_{\sum d_i=d}\prod_{i=1}^{n}{p_i^{2d_i}\over d_i!}\<\tau_{d_1}\dots
\tau_{d_n}\>_{g}.
\label{index}
\eea

One important note is in order. It is a theorem by Kontsevich that these
integrations extend continuously to the closure of the moduli space $\Mc$
following the procedure by Deligne and Mumford \cite{Mum83}.
(It means that we deal with a
stable cohomological class of curves.)

In order to compare with a matrix model we should take the Laplace
transform
over variables $p_i$ of volumes of fibers of $\pi$:
\be
\int_{0}^{\infty}dp_i\e ^{-p_i\l_i}p_i^{2d_i}=(2d_i)!\l _i^{-2d_i-1}
\ee
for the quantities standing on the right--hand side of (\ref{index}). On
the left--hand side we have
\be
\int_{0}^{\infty}\dots \int_{0}^{\infty}dp_1 \wedge\dots\wedge dp_n \e
^{-\sum p_i \l_i}\int _{\Mc }\e^{\Omega},
\ee
and due to cancellations of all $p_i^2$ multipliers with $p_i$'s in
denominators of the form $\Omega$ we get:
\be
\e ^{\Omega} dp_1\wedge\dots\wedge dp_n = \rho \prod_{e\in X_1}
dl_{e}.
\label{volume}
\ee
We use standard notations: $X_q$ is a total number of $q$--dimensional
cells
of a simplicial complex. ($X_1$ is the number of edges, $X_0$ -- the number
of vertices, etc).
$\rho$ is a positive function defined on open cells, it is equal to the
ratio of measures:
\be
\rho=\left(\prod_{i=1}^n|dp_i|\times {\Omega^d\over d!}\right):
\prod_{e\in X_1}|dl(e)|.
\label{rho}
\ee

Surprisingly, the constant $\rho$ does depend only on Euler characteristic
of the graph $\Gamma$, $\rho=2^{-\kappa}$,
\be
\rho=2^{d+\# X_1-\# X_0}.
\label{rho1}
\ee

The integral
\be
I_g(\l_{\ast}):=\int_{\Mcomb}\exp\bigl(-\sum \l_i p_i\bigr)\prod_{e\in X_1}
|dl(e)|
\ee
is equal to the sum of integrals over all open strata in $\Mcomb$. These
open
strata are in correspondence with a complete set of three--valent graphs
contributing to this order in $g$ and $n$. It is necessary also to take
into
account internal automorphisms of the graph (their number, in fact, counts
how
many replics of moduli space one may find in this cell if one treat all
$l_e$
independently). The last step is to perform the sum $\sum \l_i p_i$ in a
form dependent on $l_e$. By a standard procedure for simplicial complexes
\be
\sum_{i=1}^n \l_i p_i=\sum_{e\in X_1} l_e(\l_e^{(1)}+\l_e^{(2)}).
\ee
Here $\l_{e}^{(1)}$ and $\l_{e}^{(2)}$ are variables of two
cycles divided by $e$th edge. Performing now the Laplace transform we get
the
relation (\ref{Aut}). The quantity standing in the r.h.s. is nothing but a
term from $1/N$ expansion of the Kontsevich matrix model and eventually we
have:
\bea
&{}&\sum_{{g=0\atop n=1}}^{\infty} N^{2-2g}\alpha^{2-2g-n}
\sum_{s_1+2s_2+\dots+ks_k=d}\<(\tau_0)^{s_0}\dots (\tau_k)^{s_k}\>_g
{1\over s_0!\dots s_k!}\prod_{i=1}^n\tr {(2d_i-1)!!\over \Lambda^{2d_i+1}}
\nonumber\\
&=&\log {\int_{N\times N}DX\exp\left\{N\alpha
\tr\biggl(-\frac{X^2\Lambda}{2}
+\frac{X^3}{6}\biggr)\right\} \over
\int_{N\times N}DX\exp\left\{N\alpha \tr\biggl(-\frac{X^2\Lambda}{2}
\biggr)\right\} }
\label{MMK}
\eea
Thus the Kontsevich matrix model is a generating function for intersection
indices of the first Chern classes on moduli (orbi)spaces.

\newsection{The Penner--Kontsevich model.}

\redefine\l{\mu}

Now let us turn to the case of the Penner--Kontsevich model (PK model)
\cite{CM92a}, \cite{CM92b} with
the partition function given by
\be
{\cal Z}[\L ]={\int DX \exp \left(\alpha N \tr \left\{
-\frac 14 \L X\L X - \frac 12 \bigl[\log (1-X)+X\bigr]\right\}\right)
\over
\int DX \exp \left(\alpha N \tr \left\{
-\frac 14 \L X\L X + \frac 14 X^2\right\}\right) },\quad \L=\hbox{diag}
(\l_1,\dots ,\l_N).
\label{KPM}
\ee
It includes
in variance with the Kontsevich model all powers of $X^n$ in the potential
since it describes the partition of moduli space into cells of a simplicial
complex, the sum running over all simplices with different dimensions. (On
the language of the Kontsevich model the lower the dimension, the more and
more edges of the fat graph are reduced).

We find the Feynman rules for the Kontsevich--Penner
theory (\ref{PK}). First, as in the standard Penner model, we have
vertices of all orders in $X$. Due to rotational symmetry, the factor $1/n$
standing with each $X^n$ cancels, and only symmetrical factor
$1/\# \hbox{Aut}\,\Gamma$
survives. Also there is a factor $(\alpha/2)$ standing with each vertex. As
in the
Kontsevich model, there are variables $\l_i$ associated with each cycle.
But the form of propagator changes --- instead of $2/(\lambda_i+\lambda_j)$ we
have
$2/(\l_i\l_j+\alpha)$.

Let us consider the same case ($g=0$, $n=3$) as for Kontsevich model. One
additional diagram resulting from vertex $X^4$ arises (Fig.2).

\phantom{slava KPSS}

\vspace{4cm}

%
%
\begin{picture}(280,2)(-15,5)
\put(40,40){\oval(40,40)[b]}
\put(40,40){\oval(30,30)[b]}
\put(40,45){\oval(40,40)[t]}
\put(40,45){\oval(30,30)[t]}
\put(20,40){\line(0,1){5}}
\put(60,40){\line(0,1){5}}
\put(25,40){\line(1,0){30}}
\put(25,45){\line(1,0){30}}
\put(-10,40){\makebox{$\alpha^{-1}/3$}}
\put(20,65){\makebox{$\lambda_1$}}
\put(35,50){\makebox{$\lambda_2$}}
\put(35,30){\makebox{$\lambda_3$}}
\put(63,40){\makebox{$+\alpha^{-1}$}}
\put(100,40){\oval(30,30)[b]}
\put(100,40){\oval(20,20)[b]}
\put(100,45){\oval(30,30)[t]}
\put(100,45){\oval(20,20)[t]}
\put(85,40){\line(0,1){5}}
\put(90,40){\line(0,1){5}}
\put(110,40){\line(0,1){5}}
\put(115,40){\line(1,0){20}}
\put(115,45){\line(1,0){20}}
\put(150,40){\oval(30,30)[b]}
\put(150,40){\oval(20,20)[b]}
\put(150,45){\oval(30,30)[t]}
\put(150,45){\oval(20,20)[t]}
\put(165,40){\line(0,1){5}}
\put(160,40){\line(0,1){5}}
\put(140,40){\line(0,1){5}}
\put(120,55){\makebox{$\lambda_1$}}
\put(95,40){\makebox{$\lambda_2$}}
\put(145,40){\makebox{$\lambda_3$}}
\put(170,40){\makebox{$+\alpha^{-1}$}}
\put(205,40){\oval(30,30)[b]}
\put(205,40){\oval(20,20)[b]}
\put(205,45){\oval(30,30)[t]}
\put(205,45){\oval(20,20)[t]}
\put(190,40){\line(0,1){5}}
\put(195,40){\line(0,1){5}}
\put(215,40){\line(0,1){5}}
\put(235,40){\oval(30,30)[b]}
\put(235,40){\oval(20,20)[b]}
\put(235,45){\oval(30,30)[t]}
\put(235,45){\oval(20,20)[t]}
\put(250,40){\line(0,1){5}}
\put(245,40){\line(0,1){5}}
\put(225,40){\line(0,1){5}}
\put(215,60){\makebox{$\lambda_1$}}
\put(200,40){\makebox{$\lambda_2$}}
\put(230,40){\makebox{$\lambda_3$}}
\put(255,40){\makebox{+perm.}}
\end{picture}

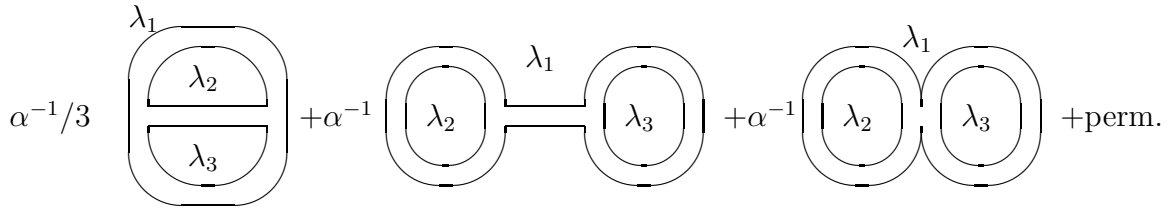
\centerline{Fig.2. g=0, s=3 contribution to Penner--Kontsevich's model.}

\vspace{6pt}

This contribution is (symmetrized over $\l_1$, $\l_2$ and $\l_3$):
\bea
&-&\frac 13\left\{{2\alpha^{-1}\over 2(\l_1\l_2-1)
(\l_1\l_3-1)}+\hbox{\ perm.\ }\right\}
+ {2\alpha^{-1}\over 6(\l_1\l_2-1)(\l_1\l_3-1)
(\l_2\l_3-1)}\nonumber\\
&+&\frac 13 \left\{{2\alpha ^{-1}\over
2(\l_1^2-1)(\l_1\l_2-1)(\l_1\l_3-1)}
+\hbox{\ perm.\ }\right\}
\label{111}
\eea
Again collecting all terms we get:
\be
{2\alpha^{-1}\over 6\prod_{i<j}(\l_i\l_j-1)}\left\{\sum_{i<j} \l_i\l_j
-2+\left({\l_2\l_3-1\over \l_1^2-1}+
{\l_1\l_2-1\over \l_3^2-1}+
{\l_1\l_3-1\over \l_2^2-1}\right)\right\},
\label{112}
\ee
and after a little algebra we obtain an answer:
\be
F_{0,3}=\alpha^{-1} {\l_1\l_2+\l_1\l_3+\l_2\l_3+1\over 3(\l_1^2-1)
(\l_2^2-1) (\l_3^2-1)}.
\label{f03}
\ee
We see that here, just as in standard Kontsevich model, the cancellation
of intertwining terms in the denominator occures that leads to
factorization
of the answer over $1/(\l_i^2-1)$--terms. This simplest example shows
that there should be some underlying geometric structures in this case as
well.

Note that technically the reason why there is the dependence only on $\tr \L^k$
$(k\le 0)$ for the partition function (\ref{KPM}) is because this model as well
as
the Kontsevich one belongs to the class of Generalized Kontsevich Models
\cite{KMMMZ}.
It means that after some simple transformation we can get from (\ref{KPM}) the
model
with the potential $\L X+V(X)$ for which only the dependence on Miwa`s times
exists.

\subsection{The Discretized Moduli Space.}

\redefine\l{\lambda}

Let us now consider a discretization of the moduli spaces $\Mc$
and $\Mcomb$:
\be
l_i\in {\bf Z}_+\cup \{0\},\quad p_i\in {\bf Z}_+, \quad\sum_{i=1}^n p_i\in
2{\bf Z}_+.
\label{d1}
\ee
So all $l_i$ and $p_i$ are now integers, but some of $l_i$ can be zeros
while all perimeters are strictly positive, the sum of all perimeters must
be even, because each edge contributes twice into this sum. We call this
(combinatorial) space $\Mcdisc$. It is worth to note that now we explicitly
include into play such points of the original $\Mc$ which are points of
reductions. It is easy to see from (\ref{d1}). While keeping all $p_i$
fixed
in a general case we can put a number of $l_j$ exactly equal zero. Some of
these configurations belong to interior of $\Mgn$, but not all --- it means
that among the points of $\Mcdisc$ there are points which lie on the
boundary $\partial \Mgn$ and they correspond to reductions of the
algebraic curve.
Also we shall use the notation $\Mdisc$ for
such subset of $\Mcdisc$ where all points of reduction are excluded.

It appears that
this choice for d.m.s. is rather natural since all the quantities
(\ref{k2}-\ref{omega}) have corresponding conterplates in this discrete
case.

First we need to define the action of the external derivative $d$ and the
integration over these (orbi)spaces. We shall write its action on
functions,
the extrapolation to the space of skewsymmetric forms is obvious
\be
df(l_1,\dots,l_k)=\sum_{i=1}^k\bigl(f(l_1,\dots,l_i+1,\dots, l_k)-
f(l_1,\dots,l_k)\bigr)dl_i
\label{d2}
\ee
As for the integral over a domain $\Omega$, there is again a proper
generalization of it to this discrete case:
\be
\int_{\Omega}f(l_1,\dots,l_k)dl_1\dots dl_k:=\sum_{{l_i\in {\bf Z}_+\cup
\{0\}
\atop \{l_1,\dots,l_k\}\in\Omega}}f(l_1,\dots,l_k).
\label{dint}
\ee

Instead of $BU(1)^{comb}$ we have (orbi)space of equivalence classes of all
sequences of non-negative integers $l_1,\dots,l_k$ modulo cyclic
permutations.
An analog of $S^1$--bundle is now a kind of ${\bf Z}_p$--``bundle'' over
this new discrete orbispace whose total space $E{\bf Z}_p^{comb}$ is an
ordinary rectangular lattice. The fiber of the bundle over the
equivalence class of
sequences $l_1,\dots, l_k$ is again the polygon with integer
lengths of edges
$l_1\dots, l_k$.

We denote $\phi_i$ coordinates on $E{\bf Z}_p$ just as in (\ref{k1}):
\be
l_i=\phi_{i+1}-\phi_i\quad (i=1,\dots,k-1),\quad l_k=p+\phi_1-\phi_k.
\label{d3}
\ee
Due to linearity of (\ref{k2}) in $l_i$, and $\phi_j$ it can be
straightforwardly generalized to our case.
Denote by ${\t \alpha}$ the 1--form on $E{\bf Z}_p^{comb}$ which is equal
to
\be
{\t\alpha} = \sum_{i=1}^k \frac{l_i}{p}\times \frac{d\phi_i}{p}.
\label{d4}
\ee
The integral of ${\t \alpha}$ over each fiber of
the universal bundle $E{\bf Z}_p^{comb}\to BU(1)^{comb}$ is equal $-1$. The
differential $d{\t\alpha}$ is the pullback of a 2--form $\t\omega$ on the
base
$B{\bf Z}^{comb}_p$,
\be
{\t\omega} = \sum_{1\leq i< j\leq k-1} \frac{dl_i}{p}\wedge
\frac{dl_j}{p}.
\label{d5}
\ee
Extrapolating these results to the whole discrete moduli space we obtain
that
the pullback ${\t\omega}_i$ of the form ${\t\omega}$ under the $i$th map
$\Mcdisc\to B{\bf Z}_p^{comb}$ represents the class ${\t c}_1(\LL_i)$.

Denote by ${\t\pi}: \Mcdisc\to \bigl[{\bf Z}_+^n\bigr]_{even}$ the
projection
to the space of perimeters with the restriction $\sum_i p_i\in 2\cdot {\bf
Z}_+$. Intersection indices are given again by  the formula:
\be
\Lan\tau_{d_1}\dots\tau_{d_n}\Ran=
\int_{{\t\pi}^{-1}(p_{\ast})}\prod_{i=1}^n
{\t\omega}_i^{d_i},
\label{d6}
\ee
where $p_{\ast}=(p_1,\dots, p_n)$ is an arbitrary sequence of positive
integer numbers with even sum and the $\pi^{-1}(p_{\ast})$ is an analogue
of
the fiber of $\Mc$ in $\Mcdisc$.

One important note is in order. These fibers $\pi^{-1}(p_{\ast})$ contain
finite number of points each  and are not isomorphic to each other. But
they are just the analogues of the
initial moduli space $\Mc$ labelled by different perimeters. This is a
point
why we call them discretized moduli spaces. It appears that
without any reference how many points from the initial $\Mc$ participate in
a fiber $\pi^{-1}(p_{\ast})$ (it can be even only one point of reduction,
as
we shall see for ${\overline {\cal M}}_{1,1}$), the relation (\ref{d6})
remains valid. Intuitive considerations which can convince us in it are
the following: Values of these intersection indices are some rational
numbers due to the orbifold nature of the initial moduli space $\Mc$. On
the language of graphs this nature reveals itself as symmetries of the
graphs. But these symmetry properties are the same whatever case --
continuum or discrete, and whatever values of perimeters we choose. Thus
preserving symmetry properties we preserve the values of cohomological
classes on both continuum and discrete moduli spaces.

\subsection{Matrix Integral for Discretized Moduli Space.}
We denote by $\t\Omega$ the two--form on $\Mcdisc$:
\be
\t\Omega = \sum_{i=1}^{n} p_i^2 \t\omega_i,
\label{dom}
\ee
which restriction to the fibers of $\t\pi$ has constant coefficients in the
coordinates $\bigl(l(e)\bigr)$. $d$ is again the complex dimension of
$\Mgn$, $d=3g-3+n$. The volume of the fiber of $\t\pi$ with respect to
$\t\Omega$
is
\bea
\hbox{vol}\bigl(\t\pi^{-1}(p_1\dots ,p_n)\bigr) &=&
\int_{\t\pi^{-1}(p_{\ast})} {\t\Omega ^d\over d!} = \frac{1}{d!}
\int_{\t\pi^{-1}(p_{\ast})}\bigl(p_1^2\t\omega_1+\dots +
p_n^2\t\omega_n\bigr)^{d}=\nonumber\\
&=&\sum_{\sum d_i=d}\prod_{i=1}^{n}{p_i^{2d_i}\over
d_i!}\Lan\tau_{d_1}\dots
\tau_{d_n}\Ran_{g}.
\label{dind}
\eea

Next step is to do a Laplace transform in both sides of (\ref{dind}). Of
course, now we should replace continuum Laplace transform by the discrete
one and also explicitly take into account that sum of all $p_i$ is even. On
the R.H.S. we have:
\be
\sum_{{p_i\in{\bf Z}_+,\atop \sum p_i\in 2{\bf Z}_+}}\e ^{-\sum_ip_i\l_i}
p_1^{2d_1}\dots p_n^{2d_n}=\prod_{i=1}^n\left(\dd{\l_i}\right)^{2d_i}\times
\frac 12 \left\{\prod_{i=1}^n\frac{1}{\e^{\l_i p_i}-1}+(-1)^n
\prod_{i=1}^n\frac{1}{\e^{\l_i p_i}+1}\right\}.
\label{dlapl}
\ee
On the L.H.S. of (\ref{dind}) we again substitute
\be
\e^{\t\Omega}\bigl.dp_1\wedge\dots dp_n\bigr|_{\sum p_i\in 2{\bf Z}_+}
=\t\rho\prod_{e\in X_1}dl_e.
\label{dmeasure}
\ee
Here the constant $\t\rho$ is the ratio of measures similar to
(\ref{rho}) and we only need to take into account the restriction that
the sum of all $p_i$ is even. It leads to renormalization of the $\rho$
for the case of d.m.s.:
\be
\t\rho =\rho/2,
\label{drho}
\ee
where $\rho$ is given by (\ref{rho1}).

Now we want to find a matrix model description for these ``new''
intersection indices. Here we immediately encounter some troubles.
Let us look which graphs correspond to different points of
$\tpip$. First, there are points of a general position for which all
$l_i$ are greater than zero, which correspond to graph with only
trivalent vertices. Second, there are such points of $\tpip$ that some
$l_i$ are zeros, but these points still do not correspond to
reductions. For example, see Fig.3, where for the torus case Fig.3a
represents a point of a general position, $l_i>0,\ i=1,2,3$,
and Fig.3b gives an example of the graph for which one (and only one) of
$l_i$ is zero. Such graphs by no means correspond to a reduction point,
but rather belong to subspace of $\Mgn$ (in Teichm\"uller
parametrization) with modular parameter $\tau$ being purely imaginary.
But certainly, if we want to include such graphs into consideration we
should consider not only trivalent vertices, but vertices of arbitrary
order. In the continuum limit we did not take
into account such graphs since they correspond to subdomains of lower
dimensions in the interior of the moduli space,
the integration measure
being continuous, and we may neglect them. Now the situation changed and we
should
take into account all such diagrams as well.

\phantom{slava KP-ss}

\vspace{4.5cm}

\begin{picture}(0,0)(-30,0)
\put(30,40){\oval(40,40)[b]}
\put(30,40){\oval(35,35)[b]}
\put(10,42.5){\line(0,-1){2.5}}
\put(50,42.5){\line(0,-1){2.5}}
\put(47.5,57.5){\oval(35,35)[bl]}
\put(47.5,57.5){\oval(30,30)[bl]}
\put(32.5,42.5){\oval(35,35)[tr]}
\put(32.5,42.5){\oval(30,30)[tr]}
\put(30,60){\line(0,-1){2.5}}
\put(32.5,60){\line(0,-1){2.5}}
\put(10,60){\line(0,-1){17.5}}
\put(12.5,60){\line(0,-1){17.5}}
\put(21.25,60){\oval(22.5,22.5)[t]}
\put(21.25,60){\oval(17.5,17.5)[t]}
\put(12.5,50){\oval(20,20)[br]}
\put(12.5,50){\oval(15,15)[br]}
\put(30,50){\oval(20,20)[tl]}
\put(30,50){\oval(15,15)[tl]}
\put(20,75){\makebox{$l_1$}}
\put(42,63){\makebox{$l_2$}}
\put(30,27){\makebox{$l_3$}}
\put(70,40){\vector(1,0){10}}
\put(65,45){\makebox{$l_3=0$}}
\put(125,35){\oval(30,30)[b]}
\put(125,35){\oval(25,25)[b]}
\put(125,45){\oval(30,30)[t]}
\put(125,45){\oval(25,25)[t]}
\put(110,42.5){\line(0,-1){7.5}}
\put(112.5,42.5){\line(0,-1){7.5}}
\put(137.5,45){\line(0,-1){10}}
\put(140,45){\line(0,-1){10}}
\put(110,40){\oval(10,10)[l]}
\put(110,40){\oval(5,5)[l]}
\put(120,40){\oval(10,10)[r]}
\put(120,40){\oval(5,5)[r]}
\put(112.5,45){\line(1,0){7.5}}
\put(112.5,42.5){\line(1,0){7.5}}
\put(112.5,37.5){\line(1,0){7.5}}
\put(112.5,35){\line(1,0){7.5}}
\put(100,45){\makebox{$l_2$}}
\put(120,65){\makebox{$l_1$}}
\put(160,40){\vector(1,0){10}}
\put(155,45){\makebox{$l_2=0$}}
\put(205,35){\oval(30,30)[b]}
\put(205,35){\oval(25,25)[b]}
\put(205,45){\oval(30,30)[t]}
\put(205,45){\oval(25,25)[t]}
\put(217.5,45){\line(0,-1){10}}
\put(220,45){\line(0,-1){10}}
\put(191.75,35){\circle*{3}}
\put(191.75,45){\circle*{3 }}
\put(225,45){\makebox{$l_1$}}
\end{picture}

\centerline{{}\hfil{Fig.3a}\hfil{}\hfil{\phantom{Figure3b}Fig.3b}\hfil{}\hfil
{\phantom{Fuga}Fig.3c}\hfil{}}

\vspace{4pt}

\centerline{Fig.3. Diagrams for different regions of $\Mcpar{1,1}$}

\vspace{6pt}

But in each $\tpip$ there are always (except the case $\Mpar{0,3}$)
genuine points of reduction (see, for example Fig.3c when two of $l_i$
are zeros). We are not able to give a matrix model description to such
points. At first sight it would mean that all the construction fails
since we still not touch the question how to ``exclude'' such reduction
points from $\tpip$ modifying in a way the relation (\ref{d6}). We shall
call $\Mdisc$ such subset of $\Mcdisc$ where all points of reduction are
excluded. Thus we
need to release somehow the integration over open $\Mgn$ from the total
integration over $\Mc$. In order to do it we shall use a stratification
procedure
by Deligne and Mumford \cite{Mum83}. The idea is to present the open moduli
space $\Mgn$ as a combination of $\Mc$ and $\Mcpar{g_j,n_j}$ of lower
dimensions. The description of this procedure in the case of modulus one
can find in \cite{Wit2}, \cite{Dij91}.

The geometrical meaning of the reduction procedure is that we
subsequently pinch handles of the surface (Fig.4). One can see that
there are two types of such reduction: in the first one when pinching a handle
we
result in the surface of genus lower by one and two additional
punctures. Thus from the space $\Mc$ we get after this reduction type
$\Mcpar{g-1,n+2}$ (Fig.4a).
In the second type
pinching an intermediate cylinder we get two surfaces of the
same total genus and two new punctures: one per each new component. It
means that the initial moduli space $\Mc$ splits into the product
$\Mcpar{g_1,n_1+1}\otimes \Mcpar{g_2,n_2+1}$, $g_1+g_2=g$, $n_1+n_2=n$
(Fig.4b).

\phantom{slava nam}

\vspace{5.5cm}

\begin{picture}(0,0)(-20,0)
\put(60,50){\oval(60,40)}
\put(60,55){\oval(40,20)}
\multiput(45,35)(15,0){3}{\circle*{3}}
\multiput(45,35)(0,-5){3}{\line(0,-1){3}}
\multiput(60,35)(0,-5){3}{\line(0,-1){3}}
\multiput(75,35)(0,-5){3}{\line(0,-1){3}}
\put(50,85){\makebox{pinching}}
\put(95,50){\vector(1,0){10}}
\put(60,75){\vector(0,-1){5}}
\put(60,60){\vector(0,1){5}}
\put(140,50){\oval(60,40)[b]}
\put(110,60){\line(0,-1){10}}
\put(170,60){\line(0,-1){10}}
\put(120,60){\oval(20,20)[t]}
\put(160,60){\oval(20,20)[t]}
\put(140,60){\oval(20,10)[b]}
\put(123,63){\circle*{3}}
\put(157,63){\circle*{3}}
\multiput(123,63)(7,14){2}{\line(1,2){5}}
\multiput(157,63)(-7,14){2}{\line(-1,2){5}}
\multiput(125,35)(15,0){3}{\circle*{3}}
\multiput(125,35)(0,-5){3}{\line(0,-1){3}}
\multiput(140,35)(0,-5){3}{\line(0,-1){3}}
\multiput(155,35)(0,-5){3}{\line(0,-1){3}}
\put(140,90){\makebox{$\tau_0$}}
\end{picture}

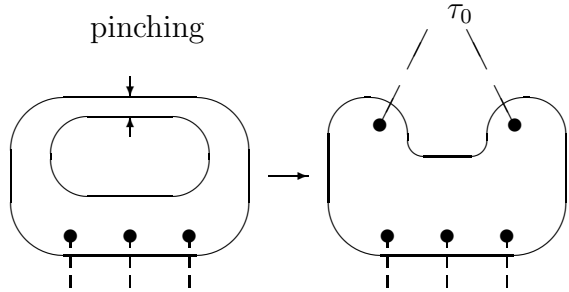
\centerline{Fig.4a. One--component type of the reduction.}

\phantom{Slava tzariy Borisy!}

\vspace{3.5cm}

\begin{picture}(0,0)(-30,0)
\put(20,35){\oval(20,20)[t]}
\put(20,30){\oval(20,20)[b]}
\put(10,35){\line(0,-1){5}}
\put(30,35){\line(1,0){30}}
\put(30,30){\line(1,0){30}}
\put(70,35){\oval(20,20)[t]}
\put(70,30){\oval(20,20)[b]}
\put(80,35){\line(0,-1){5}}
\put(70,32.5){\oval(10,15)}
\multiput(15,40)(0,-15){2}{\circle*{3}}
\multiput(15,40)(-5,0){3}{\line(-1,0){3}}
\multiput(15,25)(-5,0){3}{\line(-1,0){3}}
\put(35,50){\makebox{pinching}}
\put(90,32.5){\vector(1,0){10}}
\put(45,40){\vector(0,-1){5}}
\put(45,25){\vector(0,1){5}}
\put(125,32.5){\oval(30,30)}
\put(175,32.5){\oval(30,30)}
\put(175,32.5){\oval(15,15)}
\multiput(120,40)(0,-15){2}{\circle*{3}}
\multiput(120,40)(-5,0){3}{\line(-1,0){3}}
\multiput(120,25)(-5,0){3}{\line(-1,0){3}}
\multiput(135,32.5)(30,0){2}{\circle*{3}}
\multiput(135,32.5)(11,0){3}{\line(1,0){8}}
\end{picture}

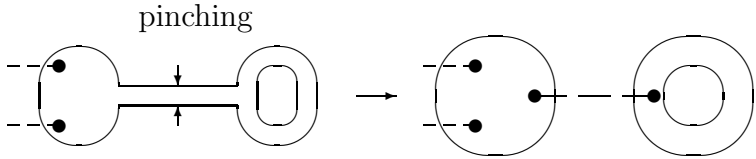
\centerline{Fig.4b. Two--component type of the reduction.}

\vspace{8pt}

One may easily check that total complex dimension of resulting spaces in
both cases is $d-1$, where $d=3g-3+n$ is the dimension of $\Mc$. So the
general receipt how to express $\Mgn$ via closed moduli spaces is to
construct an alternative sum over reductions:
\be
\Mgn = \sum_{{reductions\atop r_q=0}}^{3g-3+n}(-1)^{r_q}
{\mathop{\otimes}}_{j=1}^q\Mcpar{g_j,n_j+k_j},
\label{strat}
\ee
where sum runs over all $q$--component reductions, $r_q$ being the
reduction degree and $k_j$ being the number of the
additional punctures due to reductions.
The dimension of $\Mcpar{g_j,n_j+k_j}$ is $d_j=3g_j-3+n_j+k_j$,
\be
\sum_{j=1}^q n_j=n,\quad \sum_{j=1}^q d_j=d-r_q.
\ee
Thus we have:
\bea
&{}&\int_{\Mdisc}\e^{\t\Omega}\times \e^{\sum_i \l_i
p_i}dp_1\wedge\dots\wedge dp_n =
\frac {1}{d!}\int_{\Mcdisc} \left(\sum_{i=1}^np_i^2\t\omega_i\right)^d
\e^{\sum_i \l_i p_i}dp_1\wedge\dots\wedge dp_n +\nonumber\\
&{}&\phantom{aaa}+\sum_{{reductions\atop r_q=1}}^{3g-3+n}(-1)^{r_q}
\mathop{\otimes}\limits_{j=1}^q \int_{\Mcpar{g_j,n_j+k_j}}
\left(\sum_{a=1}^{n_j}p_a^2\t\omega_{a}\right)^{d_j}\e^{\sum_i\l_i p_i}
dp_{1}\wedge\dots\wedge dp_{n_j}.
\label{dintegral}
\eea

Now we can find using (\ref{dmeasure}) a matrix model description for
the L.H.S. of (\ref{dintegral}). Just as in the continuum case we have:
\be
\hbox{L.H.S.}= \int_{\Mdisc}\exp \left\{-\sum_{e\in X_1}l_e(\l_e^{(1)}+
\l_e^{(2)}) \right\}\times\t\rho\times\prod_{e\in X_1}|dl(e)|,
\label{lhs1}
\ee
where $\t\rho=2^{d+\# X_1-\# X_0-1}$.

This last expression can be presented as a sum over ``fat graphs''
$\Gamma$ with vertices of all valencies which are possible for given
genus and number of faces. We should again take into account the volume
of the authomorphism group for each graph which coincides with the
number of copies of equivalent domains of the moduli space $\Mgn$ which
constitute this cell of the combinatorial simplicial complex. The last
step is to do the ``integration'' over each $l(e)$ which is given by the sum
over all positive integer values of $l(e)$ (because we already took
into account all zero values of $l(e)$ doing the sum over
all graphs). Eventually we have:
\be
\hbox{L.H.S.}=
2^{d-1}\sum_{{all\atop Graphs\ \Gamma}}\frac{1}{\#
\hbox{Aut\,}(\Gamma)}\times 2^{-\# X_0}\times
\prod_{e\in X_1}\frac{2}{\e^{\l_e^{(1)}+\l_e^{(2)}}-1}.
\label{lhs2}
\ee
It is nothing but a term from the genus expansion of the
matrix model (\ref{KPM}) with
\be
\Lambda = \hbox{diag\,}\{\e^{\l_1},\dots, \e^{\l_N}\}.
\label{ll}
\ee
Then $\log \ZZ[\L]$ has the following genus expansion:
\be
\log \ZZ[\L ]=\sum_{g=0}^{\infty}\sum_{n=1}^{\infty}(N\alpha)^{2-2g}
\alpha^{-n} w_g(\l_1,\dots, \l_n).
\label{asympt}
\ee

Let us use the relations (\ref{dlapl}) in order to express the R.H.S.
of (\ref{dintegral}) via intersection indices.
\bea
&{}&w_g(\l_1,\dots, \l_n)=\frac{1}{2^{d-1}}\sum_{{reductions\atop
q-component}}(-1)^{r_q}\prod_{j=1}^q\Biggl\{
\sum_{\sum d_{\xi}=3g_j-3+n_j+k_j}\frac{1}{n_j!}\Lan
\underbrace{\tau_{d_1}\dots
\tau_{d_{n_j}}}_{n_j}\underbrace{\tau_0\dots
\tau_0}_{k_j}\,\Ran_{g_j}\Biggr.\nonumber\\
&{}&\phantom{slava}\times
\tr\,\Biggl.\prod_{k=1}^{n_j}\left(\dd{\l_k}\right)^{2d_k}\cdot\,
\frac{1}{(d_k)!}\cdot\,\frac 12
\left(\prod_{k=1}^{n_j}\frac{1}{\e^{\l_k}-1}
+(-1)^{n_j}\prod_{k=1}^{n_j}\frac{1}{\e^{\l_k}+1}\right)\Biggr\}.
\label{dgen}
\eea
This formula is our main result.
For practical reasons it is sometimes convenient to rewrite
\bea
&{}&\sum_{\sum d_{\xi}=3g_j-3+n_j+k_j}\frac{1}{n_j!}\Lan
\overbrace{\underbrace{\tau_{d_1}\dots
\tau_{d_{n_j}}}_{n_j}\underbrace{\tau_0\dots
\tau_0}_{k_j}}^{s_j}\,\Ran_{g_j}=\nonumber\\
&{}&\phantom{Il'ichy slava}
\sum_{{b_0+b_1+\dots +b_k=n_j\atop
0\cdot b_0+1\cdot b_1+\dots +k\cdot b_k=3g_j-3+s_j}}\,
\frac{1}{b_0!\dots b_k!}\Lan (\tau_0)^{b_0}\dots (\tau_k)^{b_k}
\underbrace{\tau_0\dots \tau_0}_{k_j}\,\Ran_{g_j}.
\label{dorder}
\eea
Taking this expression for the case ${\cal M}_{0,3}$ (without reductions) and
reminding that $\<\tau_0^3\>_0^{}=1$ we immediately get the answer (\ref{f03})
after
a substitution $\mu_i=\e ^{\lambda_i}$.

Since the matrix model (\ref{KPM}) is equivalent to the hermitian
one--matrix with an arbitrary potential, then the formulae
(\ref{asympt})--(\ref{dorder}) above give the solution to such models
in geometric invariants of the d.m.s. We shall see in the next section
that the relation (\ref{i1}) holds true, i.e. the intersection indices
coincide for $\Mc$ and $\Mcdisc$ thus expressing the solution to one
matrix model beyond the double scaling limit exactly in terms of
variables which live just in the Kontsevich model or, equivalently, in
the d.s.l.!

\def\bb{\beta}
\def\Tdisc{{\cal T}^{disc}_{g,n}}

There is nevertheless some nontrivial point in the final
relation (\ref{dgen}). Namely, it is the ``sum over reductions''. This
sum appears to be rather involved by the following reasons: When we
considered the orbits $\tpip$ we assumed that they belonged to
{\sl one} copy of the moduli space $\Mc$. But when we deal with the
cell decomposition it is much more convenient first to consider the
total simplicial complex (which we shall denote $\Tgn$) and only
afterward take into account internal authomorphisms of $\Tgn$ which
eventually produce $\Mc$ as a coset over a symmetry group $G_{g,n}$.
One may
consider instead of $\pi$ and $\t\pi$ mappings $\bb$ and $\t\bb$
correspondingly, $\bb: \Tgn\otimes{\bf R}_+^n\to \Mcomb$ and
$\t\bb: \Tdisc[p_{\ast}]\times[{\bf Z}_+^n]_{even}\to \Mcdisc$, where
$\Tdisc$ are again finite (nonisomorphic) sets of points of $\Tgn$
supplied with the discrete de Rham cohomology structure.

For these spaces $\Tgn$ an analogue of the formula (\ref{dgen}) exists.
The only difference is that we should multiply all indices
$\Lan\tau_{d_1}\dots\tau_{d_n}\Ran_g$ by the order of the symmetry
group $G_{g,n}$.
We shall present arguments in the next section that there is such
choice of $\Tgn$ that all reductions ($\TT_{g_j,n_j+k_j}$) are counted
integer number of times. But this number of copies
multiplied by the order of the group
$G_{g_j,n_j+k_j}$ does not necessarily divisible by the order of
$G_{g,n}$. Thus when we write in the formula (\ref{dgen}) ``the sum
over reductions'' we should keep in mind that the coefficients in this
sum are not necessarily integers! (See Section~5 for an example).

The next section is devoted to the comparison of the matrix integrals
in the Kontsevich and matrix model for d.m.s. using exclusively matrix
model tools, that permits to prove our basic relation (\ref{i1}).

\newsection{Comparison of two matrix models}

\def\H{\eta}

This section is based on the results of the papers \cite{ACM} and
\cite{ACKM}. It was explicitly demonstrated in \cite{CM92b} \cite{KMMM}
that the matrix model (\ref{KPM}) is equivalent to the standard hermitian
one--matrix model
\be
\ZZ [g,\t{N}]=\int_{\t{N}\times \t{N}}d\phi\exp\bigl(-\t{N}\tr V(\phi)\bigr),
\label{4.1}
\ee
where the integration goes over hermitian $\t{N}\times \t N$ matrices and
\be
V(\phi)=\sum_{j=1}^{\infty}\frac{g_j}{j}\,\phi^j
\label{4.2}
\ee
is a general potential. Then the following relation holds:
\be
\ZZ [g,\t N(\aa)]=\e^{-N\tr \H^2/2}\ZZ_P[\H,N],\quad \t N(\aa)=-\aa N.
\label{4.3}
\ee
Here the partition function $\ZZ_P[\H,N]$ is
\be
\ZZ_P[\H,N]=\int_{N\times N}dX\exp\left[ N\tr\left(-\H X-\frac 12
X^2-\aa\log X\right)\right],
\label{4.4}
\ee
the integral being done over hermitian matrices of {\it another}
dimension $N\times N$ and the set of the coupling constants (\ref{4.2})
being related to the matrix $\H$ by the Miwa transformation
\be
g_k=\frac 1N \tr\H^{-k}-\delta_{k,2}\ \hbox{for}\ k\ge 1,\quad
g_0=\frac 1N \tr\log\H^{-1}.
\label{4.5}
\ee
Now after substitution
\be
\H=\sqrt{\aa}(\L+\L^{-1})
\label{4.6}
\ee
and making the change of variables $X\to (X-1)\L\sqrt{\aa}$ we
reconstruct the integral (\ref{KPM}) (with $\aa$ multiplied by two).

Note that we can do a limiting procedure (which is a sort of the double
scaling limit for the standard model (\ref{4.1})) resulting in the
Kontsevich integral (2.2) starting from Kontsevich--Penner model
(\ref{KPM}). It looks even more natural in terms of this model
than in terms of the one--matrix integral (\ref{4.1}). Namely,
let us take in (\ref{KPM})
\be
\L=\e^{\ep\l},\quad \aa=\frac{1}{\ep^3}.
\label{4.7}
\ee
Then after rescaling $X\to\ep X$ in the limit $\ep\to 0$ we explicitly
reproduce (2.2) from (\ref{KPM}). During this procedure we can keep
constant the size $N$ of matrices of (\ref{KPM}), but the size $\t N(\aa)$
of the matrices of hermitian model goes to infinity in the limit
$\ep\to\infty$. There is a question whether this limit is a genuine
d.s.l., because there exists another limiting procedure which generate a
square of the Kontsevich model as the limit of (\ref{KPM}), but for our
present purposes we shall use this simplest scaling limit and we refer to
it as d.s.l.

\subsection{Review of the solutions to Kontsevich and KP models.}

Since the models (\ref{KPM}) and (\ref{4.1}) are equivalent we can use
the explicit answers for (\ref{4.1}) found in \cite{ACKM} in order to
check the validity of our formulae (\ref{dgen}) and to compare the values
of intersection indices in both Kontsevich model (\ref{Konts}) and the model
(\ref{KPM}). Both these models were solved in genus expansion in terms of
momenta. For the Kontsevich model this solution was presented in
\cite{IZ92} and for (\ref{4.4}) or, equivalenly, (\ref{4.1}) --- in
\cite{ACKM}. Here we present the results. (Throughout this section the
expansion
parameter $\aa$ should be replaced by $-2\aa$ in order to compare with the
results of
\cite{ACKM}.

{\bf 1.} The solution to the Kontsevich model is
\be
\log\ZZ_K[N,\L]=\sum_{g=0}^{\infty}N^{2-2g}F_g^{Kont}.
\label{4k1}
\ee
For the genus expansion coefficients we have
\be
F_g^{Kont}=\sum_{{\aa_j>1\atop \sum_{j=1}^n(\aa_j-1)=3g-3}}
\<\tau_{\aa_1}\dots \tau_{\aa_n}\>_g^{}\,\frac{I_{\aa_1}\dots I_{\aa_n}}
{(I_1-1)^{\aa}}\ \ \hbox{for}\ \ g\ge 1,
\label{4k2}
\ee
where $\<\cdot\>_g$ are just intersection indices and the moments $I_k$'s
depending on an external field $M$ are defined by
\be
I_k(M)=\frac{1}{(2k-1)!!}\,
\frac{1}{N}\sum_{j=1}^N\frac{1}{(m_j^2-2u_0)^{k+1/2}}\quad k\ge 0,
\label{4k3}
\ee
and $u_0(M)$ is determined from the equation
\be
u_0=I_0(u_0,M).
\label{4k4}
\ee

{\bf 2.} The solution to the model (\ref{4.4}) can be written as
\be
\log\ZZ_P[N,\H]=\sum_{g=0}^{\infty}N^{2-2g}F_g,
\label{4p1}
\ee
where
\be
F_g=\sum_{\aa_j>1,\ \bb_i>1}\<\aa_1\dots \aa_s; \bb_1\dots
\bb_l|\aa,\bb,\g\>_g\frac{M_{\aa_1}\dots M_{\aa_s}J_{\bb_1}\dots
J_{\bb_l}}
{M_1^{\aa}J_1^{\bb}d^{\g}}\quad g>1.
\label{4p2}
\ee
This solution was originated from the one--cut solution to the loop equations
in
the hermitian one--matrix model, $x$ and $y$ being endpoints of this cut,
$d=x-y$, and for momenta $M_k$, $J_k$ we have
\bea
M_k&{}&=\frac 1N \sum_{j=1}^N\frac{1}{(\h_j-x)^{k+1/2}(\h_j-y)^{1/2}} -
\delta_{k,1}\quad k\ge 0,
\label{4p3}\\
J_k&{}&=\frac 1N \sum_{j=1}^N\frac{1}{(\h_j-x)^{1/2}(\h_j-y)^{k+1/2}} -
\delta_{k,1}\quad k\ge 0.
\label{4p4}
\eea
The brackets $\<\cdot\>_g$ denote rational numbers, the sum is finite in
each order in $g$, while the following restrictions are fulfilled: If we
denote by $N_M$ and $N_J$ the total powers of $M$'s and $J$'s
respectively, i.e.
\be
N_M=s-\aa,\quad N_J=l-\bb,
\ee
then it holds that $N_M\le 0$, $N_J\le 0$ and
\bea
F_g:&{}&\quad\quad N_M+N_J=2-2g,\nonumber\\
F_g:&{}&\sum_{i=1}^s(\aa_i-1)+\sum_{j=1}^l(\bb_j-1)+\g = 4g-4\nonumber\\
F_g:&{}&\sum_{i=1}^s(\aa_i-1)+\sum_{j=1}^l(\bb_j-1)+\g \le 3g-3
\label{4p4a}
\eea

We again have a nonlinear functional equations determining the positions
of the endpoints $x$ and $y$:
\bea
\frac 1N\sum_{i=1}^N\frac{1}{\sqrt{(\h_i-x)(\h_i-y)}}-\frac{x+y}{2} &=&0,
\label{4p5}\\
\frac 1N\sum_{i=1}^N\frac{\h_i-\fr{x+y}{2}}{\sqrt{(\h_i-x)(\h_i-y)}}
-\frac{(x-y)^2}{8} &=&-2\aa+1.
\label{4p6}
\eea

The solutions to the first two genera have as usual some peculiarities.
For $g=1$ we have
\be
F_1=-\frac{1}{24}\log M_1J_1d^4,
\label{4p7}
\ee
and for zero genus we have after taking a double derivative in $\aa$ in
order to exclude divergent parts:
\be
\frac{\hbox{d}^2}{\hbox{d}\aa^2}F_0= 4\log d.
\label{4p8}
\ee
The last property of the expression (\ref{4p2}) which we want to notice
here is its symmetry under interchanging $x$ and $y$, or equivalently,
$M_i$ and $J_i$:
\be
\<\aa_1\dots \aa_s; \bb_1\dots \bb_l|\aa,\bb,\g\>_g=(-1)^{\g}
\<\bb_1\dots \bb_l; \aa_1\dots \aa_s|\bb,\aa,\g\>_g.
\label{4p9}
\ee
As we shall see in a moment this symmetrical relation is in a direct
connection with the symmetrization $\e^{\l}\to -\e^{\l}$ in the formula
(\ref{dgen}).

{\bf 3.} In the d.s.l. $\ep\to 0$ we may put
\be
y=-\frac{\sqrt{2}}{\ep^{3/2}},\quad
x=\frac{\sqrt{2}}{\ep^{3/2}}+\sqrt{2}\,u_0+\dots,
\label{4d1}
\ee
and the equation (\ref{4k4}) arises. The scaling behaviour of the momenta
$M_k$, $J_k$ and $d$ is
\bea
&{}&J_k\to -2^{-(3k/2+1)}\ep^{(3k+1)/2}I_0+\delta_{k1},\nonumber\\
&{}&M_k\to
-2^{(k-1)/2}\ep^{-(k-1)/2}((2k-1)!!I_k-\delta_{k1}),\nonumber\\
&{}&d\to 2^{3/2}\ep^{-3/2}
\label{4d2}
\eea
Thus only the terms without $J_k$--dependence and only of the highest
order in $\aa_i$ survive in the d.s.l. in which the expression
(\ref{4p2}) is reduced to the answer for the Kontsevich model
(\ref{4k2}). Then the coefficients
$\<\aa_1\dots \aa_s; \,\{\hbox{nothing}\}\,|\aa,0,\g\>_g$ coincide (up to
some factorials and powers of two) with the Kontsevich intersection
indices $\<\tau_{\aa_1}\dots \tau_{\aa_n}\>_g$. Having an explicit
solution of the form (\ref{4p2}) one may see it directly. In
\cite{ACKM} an iterative procedure was proposed in order to find
coefficients of the expansion (\ref{4p2}) and all these coefficients
were found in the genus 2 (for $g=0,1$ see \cite{ACM}). It was proved
there that coefficients of the highest order in $\aa_k$ coincide in a
proper normalization with the Kontsevich indices. It will permit us to
prove the relation (\ref{i1}) for these intersection indices in the next
subsection.

\subsection{Relation between momenta and d.m.s. variables}

\def\eL{\e ^{\lambda}}

Now our purpose is to study how one may reexpress the answers of the type
(\ref{4p2})
in terms of the quantities standing in the R.H.S. of (\ref{dgen}). It is rather
technical question, but important one if we want to deal with concrete
expresions.

At first, let us expand both momenta $M_k$, $J_k$ and the restriction equations
(\ref{4p5},~\ref{4p6}) in terms of $\l$--variables, where
$\eta=\sqrt{\aa}(\eL+\e^{-\l})$. Then for the endpoints of the cut we have:
\be
x=2\sqrt{\aa}+\xi,\quad y=-2\sqrt{\aa}+\beta,
\label{h1}
\ee
where $\xi$ and $\beta$ themselves are some polynomials in the higher momenta
$M_i$
and $J_i$ with $i,j\ge 0$. Thus, after a little algebra we shall get for, say,
the
moment $M_k$:
\be
M_k=\frac 1N\tr \frac{(\eL)^{k+1}}{\sqrt{\aa}
\Bigl((\eL-1)^2-\frac{\xi}{\sqrt{\aa}}\eL\Bigr)^{k+1/2}
\Bigl((\eL+1)^2-\frac{\beta}{\sqrt{\aa}}\eL\Bigr)^{1/2}}-\delta_{k,1}.
\label{h2}
\ee
(For $J_k$ the expression is just the same with interchanging the powers
$k+1/2$ and
$1/2$ for the two terms in the denominator.

It is convenient now to introduce new momenta:
\bea
\t M_k&=&\frac 1N\tr \frac{\sqrt{\eta-y}}{(\eta-x)^{k+1/2}},\nonumber\\
\t J_k&=&\frac 1N\tr \frac{\sqrt{\eta-x}}{(\eta-y)^{k+1/2}},
\label{h3}
\eea
that are related to the initial ones by the following relations:
\bea
\t M_k&=&M_{k-1}+\delta_{k,2}+d(M_k+\delta_{k,1}),\nonumber\\
\t J_k&=&J_{k-1}+\delta_{k,2}-d(J_k+\delta_{k,1}),\nonumber\\
M_0=J_0&=&(\t M_0-\t J_0)/d.
\label{h4}
\eea
Then for these new $\t M_k$ we have
\be
\t M_k=\frac 1N\tr \frac{1}{\sqrt{\aa}^k}\,\frac{(\eL +1)\e^{\l k}}{(\eL
-1)^{2k+1}}
\frac{\biggl[1-\frac{\beta}{\sqrt{\aa}}\frac{\eL}{(\eL +1)^2}\biggr]^{1/2}}
{\biggl[1-\frac{\xi}{\sqrt{\aa}}\frac{\eL}{(\eL -1)^2}\biggr]^{k+1/2}}.
\label{h5}
\ee
The expansion in (\ref{h5}) goes over the terms
\be
H_{ab}=\frac 1N\tr \frac{(\eL+1)\e^{a\l}}{(\eL-1)^{2a+1}}\cdot\frac{\e^{b\l}}
{(\eL+1)^{2b}},
\label{h6}
\ee
where $b\ge 0$, $a\ge k$.

\def\pp{\partial}

Let us prove now that $H_{ab}$ can be presented as a linear sum of
\bea
L_a&=&\frac 1N\tr\frac{\partial^{2a}}{\partial\l
^{2a}}\,\frac{1}{\eL-1},\nonumber\\
R_b&=&\frac 1N\tr\frac{\partial^{2b}}{\partial\l ^{2b}}\,\frac{1}{\eL+1},
\label{h7}
\eea
i.e. the sum goes over only even powers of derivatives in $\l$:
\be
H_{ab}=\sum_{i=0}^a\aa^i_{ab}L_i+\sum_{j=0}^{b-1}\beta^j_{ab}R_j.
\label{h8}
\ee
We begin with considering the case $b=0$. Then it is easy to see after a little
algebra
that
\be
\frac{\pp^2}{\pp\l^2}H_{t0}=t^2H_{t0}+2(t+1)(2t+1)H_{t+1,0}.
\label{h9}
\ee
It is enough to prove now that $H_{t1}$ can be presented in the form
(\ref{h8}). At
first,
\be
H_{01}=\frac 1N\tr\frac{\eL}{(\eL-1)(\eL+1)} =\frac 12(L_0+R_0),
\ee
and it is trivial to see that
\be
H_{t0}=H_{t-1,1}+4H_{t,1}.
\label{h11}
\ee
Thus $H_{t1}$ can be always written in the form (\ref{h8}). Now by induction,
multiplying both sides on $\frac{\eL}{(\eL+1)^2}$ and using again the relation
(\ref{h8}) we shall prove it for an arbitrary $H_{tb}$.

Let us keep now only terms of zero and first orders in traces of $\l$ in the
expressions for momenta. Then we get:
\bea
M_k&{}&\sim \frac{1}{\sqrt{\aa}^{k+1}}\frac 1N \tr\frac{\e^{\l(k+1)}}
{(\eL-1)^{2k+1}(\eL+1)}+\delta_{k1},\nonumber\\
J_k&{}&\sim \frac{1}{\sqrt{\aa}^{k+1}}\frac 1N \tr\frac{\e^{\l(k+1)}}
{(\eL-1)(\eL+1)^{2k+1}}+\delta_{k1},\nonumber\\
d&{}&\sim \sqrt{\aa}\left\{4-\frac{1}{\aa}\cdot\frac 1N \tr\frac{2}
{(\eL-1)(\eL+1)}\right\}.
\label{h12}
\eea
If we pay an attention to the terms surviving in the d.s.l. then these terms
are just
the ones arising from the term without reductions on the L.H.S. of
(\ref{dgen}). Then
we eventually prove our basic relation (\ref{i1})
\be
\Lan \tau_{d_1}\dots\tau_{d_n}\Ran_g = \< \tau_{d_1}\dots\tau_{d_n}\>_g.
\label{h13}
\ee

Note that $L_a$ and $R_a$ are just analogues of the Kontsevich's times
$T_n=(2n-1)!!
\L ^{2n+1}$. They are transforming into $T_n\cdot 2^n(n-1)!$ in the d.s.l. and
in both
cases there is no dependence on odd derivatives in $\L$. As we have mentioned
in
Introduction there are two natural ways to do d.s.l. in the KP model. While
doing the
one we already considered (\ref{4d1}--\ref{4d2}) only one set of times
$\{L_n\}$
survived and it is going to the set $\{T_n\}$ in the limit $\ep\to 0$. But if
we choose
$\L$ to be symmetrical, $\L=\hbox{diag}\{\l_1,\dots,\l_{N/2},-\l_1,
\dots,-\l_{N/2}\}$,
then another limit is possible when $L_i=R_i$ and each of these sets generates
$\{T_n\}$ thus giving a square of the integral (\ref{Konts}).

There is one question we want to discuss now. Let us consider the sum over
multicomponent reductions on the L.H.S. of (\ref{dgen}). Using the matrix model
technique we have an opportunity to distinguish between diferent types of
reductions
mostly due to that remarkable fact that symmetrization $\eL\to -\eL$ goes in
each
component separately. Only this property makes $\l$--dependent terms different
for, say, $\<\tau_1(\tau_0)^3\>_0\cdot\<\tau_2\tau_0\>_1$ and
$\<\tau_2\tau_1(\tau_0)^4\>_0$ (see Fig.5) --- both these terms appear during
the
reduction procedure of the genus two surface with two punctures. But one of
them is
due to the two--component reduction and another -- of the one--component type.
Evidently, when fixing the number of punctures, $n$, only terms containing
products of
exactly $n$ traces of $\l$ contribute to the L.H.S. of (\ref{dgen}) and
it is a technical problem to pull
these terms from the genus expansion in terms of momenta.

\phantom{Slava, slava}

\vspace{5cm}

\begin{picture}(0,0)(-30,0)
\put(60,50){\oval(70,40)}
\multiput(37.5,65)(15,0){4}{\circle*{3}}
\multiput(37.5,65)(0,5){3}{\line(0,1){3}}
\multiput(52.5,65)(0,5){3}{\line(0,1){3}}
\multiput(67.5,65)(0,5){3}{\line(0,1){3}}
\multiput(82.5,65)(0,5){3}{\line(0,1){3}}
\multiput(42.5,80)(30,0){2}{\oval(10,10)[tl]}
\multiput(47.5,80)(30,0){2}{\oval(10,10)[tr]}
\multiput(44,85)(30,0){2}{\line(1,0){2}}
\multiput(30,50)(60,0){2}{\circle*{3}}
\multiput(30,50)(-5,0){3}{\line(-1,0){3}}
\multiput(90,50)(5,0){3}{\line(1,0){3}}
\put(40,15){\makebox{$\<\tau_2\tau_1(\tau_0)^4\>_0^{}$}}
\put(145,50){\oval(30,30)}
\put(195,50){\oval(30,30)}
\put(195,50){\oval(15,15)}
\multiput(137.5,60)(15,0){2}{\circle*{3}}
\multiput(137.5,60)(0,5){3}{\line(0,1){3}}
\multiput(152.5,60)(0,5){3}{\line(0,1){3}}
\put(142.5,75){\oval(10,10)[tl]}
\put(147.5,75){\oval(10,10)[tr]}
\put(144,80){\line(1,0){2}}
\multiput(135,50)(-5,0){3}{\line(-1,0){3}}
\multiput(205,50)(5,0){3}{\line(1,0){3}}
\multiput(155,50)(30,0){2}{\circle*{3}}
\multiput(135,50)(70,0){2}{\circle*{3}}
\multiput(155,50)(11,0){3}{\line(1,0){8}}
\put(130,20){\makebox{$\<\tau_1(\tau_0)^3\>_0^{}$}}
\put(185,20){\makebox{$\<\tau_2\tau_0\>_1^{}$}}
\end{picture}

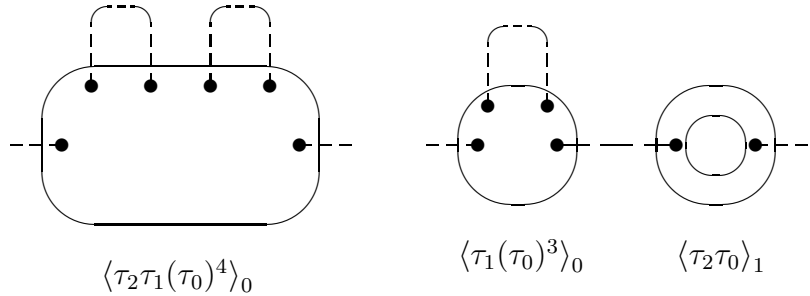
\centerline{Fig.5. Two examples of one-- and two-component reduction for ${\cal
M}_{2,2}$}

\vspace{8pt}

In the next section we apply all experience already obtained to the simplest
case of
$\Mcpar{1,1}$ -- a torus with one puncture moduli space. This consideration
enables us
to make a conclusion about the structure of any modular space.

\newsection{Orbifold structure of the moduli spaces.}

In this section we want to shed a light on the question about the
underlying structure of
d.m.s. and the moduli spaces themselves.

\subsection{The moduli space $\Mcpar{1,1}$.}
\def\MO{\Mpar{1,1}}
\def\MOC{\Mcpar{1,1}}
\def\varkappa{\kappa}
\def\Kgs{\varkappa_{g,s}}
\def\KO{\varkappa_{1,1}}

Let us consider first an example of well--known--to--everybody modular space
$\MO$,
i.e. the torus with one puncture. While thinking of it everyone immediately
imagine
the copy of this modular figure in Teichm\"uller upper half--plane, namely, a
strip
from $-1/2$ to $1/2$ along imaginary axis bounded from below by a segment of a
semicircle of the radius 1 with the origin at zero point. In order to get the
modular
space itself we should identify both sides of the strip as well as two halves
of this
segment being correspondingly on the left and on the right of the imaginary
axis $\Re
z=0$. From this picture it is clear that there are three peculiar points where
the
metric on the moduli space appears not to be conformally flat, namely, $z=i$
(square
point), $z=\e^{i\pi/3}$ (or, the same, $\e^{2i\pi/3}$) (triple point), and
$z=i\infty$
(infinity point). All these points also have a property that each of them
is stable under the action of some operator from the modular
transformation group. For triple point the subgroup of such operators has the
third
order, for square point -- it is of order 2, and for the infinity point -- of
an
infinite order.

This means that we can consider the modular space $\MO$ as an orbifold of the
(open)
upper half--plane. It was a way how Harer and Zagier \cite{HZ86} introduced
virtual
Euler characteristics for such spaces. And it was Penner \cite{Pen86} who had
found a
simple one--matrix hermitian model with the potential $\log(1+X)-X$ in order to
generate these characteristics. We shall exploit the ideology of the Penner
approach
\cite{Pen86}, \cite{DV90}
{}.
The idea was to assign factor 1 to each edge (instead of an arbitrary length as
it was
in the Kontsevich case). Then for an arbitrary $\Mgn$ there is one--to--one
correspondence between cells of the simplicial decomposition of the {\sl open}
moduli
space $\Mgn$ and graphs of the Penner model, more, the symmetrical coefficients
for each cell and corresponding graph coincide. Then the virtual Euler
characteristic
$\Kgs$ can be calculated by the formula:
\be
\Kgs=\sum_{{cells\atop (Graphs)}}\,\frac{(-1)^{n_G}}{\#\hbox{Aut}\,G},
\label{x1}
\ee
where $n_G$ is the codimension of the cell in the simplicial complex.

In the case of $\MO$ the triple point graph corresponds to the higher
dimensional cell,
square point graph -- to the cell of codimension 1. In the complex there is
also
an infinity point of the lowest dimension, but there is no graph corresponding
to
it. Thus, for the virtual Euler characteristic we get
\be
\KO=\frac 13\cdot (-1)^0+\frac 12\cdot (-1)^1+\frac {1}{\infty}\cdot
(-1)^2=-\frac 16.
\label{x2}
\ee

Let us now consider the same case, but already in the  Kontsevich
parametrization. We
know that there are three types of diagrams depictured on Fig.3a--c. The case
of Fig.3a
when all $l_i$ are greater than zero corresponds to the cell of the higher
dimension. In
$\Mcpar{1,1}$ it is a domain where $\sum_{i=1}^3 l_i=p/2$, i.e. it is an
interior of
the equilateral triangle. Note that due to two possibilities to choose an
orientation
there are two such congruent cells. The next
case is when one of $l_i$ is equal zero (Fig.3b).
When putting equal zero various $l_i$ we tend to the boundary of the previous
case --
they are open intervals lying on the boundary of the triangles. But it is not
the
whole boundary yet --- it remains one point at the summit of the triangles and
it
corresponds to the last case, Fig.3c, where two of $l_i$ are equal zero, that
gives us
a reduced case. The unique reduction of the torus with one puncture is the
sphere
with three punctures which modular space $\Mpar{0,3}$ consists from only one
point.

Now let us draw this simplicial complex ${\cal T}_{1,1}$
graphically (Fig.6). We are to indetify the
opposite edges of this parallelogram thus obtaining the torus. This torus
complex
consists from two open triangles (Fig.3a), three edges partiting these
triangles
(Fig.3b) and the unique point of reduction -- the vertex (Fig.3c). The centers
of the
triangles marked by small discs correspond to two copies of the triple point
and centers of edges -- to three copies of the square point
(small circles). We have six copies of the
original modular figure on this torus, one of them is coloured in black.

\phantom{slava}

\vspace{4.5cm}

\begin{picture}(0,0)(-80,0)
\multiput(20,20)(40,0){2}{\circle*{3}}
\multiput(40,53.33)(40,0){2}{\circle*{3}}
\multiput(20,20)(40,0){2}{\line(3,5){15}}
\multiput(40,53.33)(40,0){2}{\line(-3,-5){15}}
\put(40,53.33){\line(3,-5){15}}
\put(60,20){\line(-3,5){15}}
\multiput(20,20)(20,33.33){2}{\line(1,0){25}}
\multiput(60,20)(20,33.33){2}{\line(-1,0){25}}
\multiput(20,20)(12.5,7){5}{\line(2,1){10}}
\multiput(60,20)(-12.5,7){3}{\line(-2,1){10}}
\multiput(40,53.33)(12.5,-7){3}{\line(2,-1){10}}
\multiput(40,53.33)(0,-13){3}{\line(0,-1){10}}
\multiput(60,20)(0,13){3}{\line(0,1){10}}
\multiput(40,31.11)(20,11.11){2}{\circle*{2}}
\multiput(30,36.66)(20,0){3}{\circle{3}}
\put(40,20){\circle{3}}
\put(60,53.33){\circle{3}}
\put(40,53.33){\vector(0,1){10}}
\put(20,20){\vector(-2,-1){10}}
\put(60,20){\vector(2,-1){10}}
\put(42,33.2){\line(0,1){18}}
\put(44,34.31){\line(0,1){15.78}}
\put(46,35.42){\line(0,1){13.56}}
\put(48,36.53){\line(0,1){11.34}}
\put(50,37.64){\line(0,1){9.12}}
\put(52,38.75){\line(0,1){6.9}}
\put(54,39.86){\line(0,1){4.68}}
\put(56,40.97){\line(0,1){2.46}}
\put(45,65){\makebox{$l_1$}}
\put(20,10){\makebox{$l_2$}}
\put(75,15){\makebox{$l_3$}}
\end{picture}

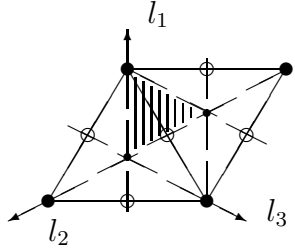
\centerline{Fig.~6. A simplicial complex for $\Mcpar{1,1}$ in the Kontsevich
picture.}

\vspace{8pt}

Thus we result in the conclusion that in the Kontsevich's parametrization the
modular
space $\MOC$ is the orbifold of a torus ${\bf T}^1$
with parameters $(1,\e^{i\pi/3})$ which
possesses an internal symmetry group $G_{1,1}$ of the sixth order. Thus
\be
\MOC={\bf T}^1/G_{1,1}.
\label{x3}
\ee
This torus is by all means totally flat compact space. And here is a point
which
is different from
the Penner construction of orbifolds of the upper half--plane, because
there we had infinity points of {\sl infinite} order, and here the order of
this point
is obviously finite! In our case it is a point of sixth order.
It means that for this case the formula (\ref{x2}) will change and using
(\ref{x1}) we
should add $1/6$ to (\ref{x2}) thus obtaining zero
for our new virtual Euler characteristic in the Kontsevich
picture (as it should be for the torus).

To complete this geometric part, note that we can think about the torus ${\bf
T}^1$ as
the fundamental domain of the subgroup $\Gamma_2$ of the modular group. This
domain is
depicted on Fig.7 and it again consists from six copies of the modular figure.
Black
discs mark the positions of triple points and small circles -- the one of
square points.
If we identify left half--line with right half--circle and vice versa we shall
obtain our torus (if we do not care about conformal properties  of this
transformation
at the infinity point).

\vspace{4.0cm}

\begin{picture}(0,0)(-80,0)
\put(15,20){\line(1,0){50}}
\put(20,20){\line(0,1){55}}
\put(60,20){\line(0,1){55}}
\multiput(30,20)(20,0){2}{\oval(20,20)[t]}
\multiput(20,60)(40,0){2}{\circle{3}}
\multiput(36,28)(8,0){2}{\circle{2}}
\put(40,31.6){\circle*{2}}
\put(40,40){\circle{3}}
\put(40,54.8){\circle*{3}}
\put(20,13){\makebox{0}}
\put(38,13){\makebox{$1/2$}}
\put(60,13){\makebox{1}}
\end{picture}

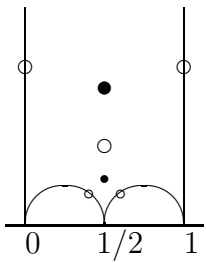
\centerline{Fig.7. The fundamental domain for subgroup $\Gamma_2$ of the
modular group.}

\vspace{8pt}

Let us turn now to our basic formula (\ref{dgen}). First, using diagram
technique for
the matrix model (\ref{KPM}) it is easy to get the answer (after substitution
$\L=\e^{\l}$). Combining all terms we obtain:
\be
F_{1,1}=\aa^{-1}\frac{3\e^{2\l}-1}{6(\e^{2\l}-1)^3_{}},
\label{x4}
\ee
and we need to express it in terms of derivatives (\ref{h7}). Note that the
formula
(\ref{x4}) one can obtain from the expansions (\ref{4p7}) and (\ref{h12})
substituting
$\aa\to -\aa/2$. After a little algebra we
get an answer:
\be
F_{1,1}=\frac{1}{48\aa}\cdot\frac{\pp^2}{\pp\l^2}\left[\frac{1}{\eL
-1}-\frac{1}{\eL+1}
\right]-\,\frac{1}{12\aa}\left[\frac{1}{\eL -1}-\frac{1}{\eL+1}\right].
\label{f11}
\ee
The first term gives us the proper value of $\<\tau_1\>_1^{}=1/24$.
As for the second term, it was originated from the ``sum over reductions'' and
the only
reduction of the torus is the sphere with three punctures for which
$\<\tau_0^3\>_0^{}=1$. We see that the sum over reductions gives an additional
fractional factor $1/6$, but now we know the nature of it.
In the simplicial complex (Fig.6) there are six copies of the modular
space $\MOC$ and only one of the infinity point.
So we see, that we just have ``one sixth'' of this
point contributing to the expression (\ref{dgen}) in this order in $g$ and $n$.

Now we are able to understand the structure of d.m.s. for $\MOC$ and starting
with this
example we shall formulate a hypothesis about the structure of an arbitrary
$\Mc$.
Remind that in the Kontsevich parametrization we use the form $\Omega$
(\ref{omega}) in order to evaluate
the volume of the corresponding modular space. The fact
that the intersection indices coincide for both continuum and discrete cases
means that
it does not matter how we calculate the total volume of the torus ${\bf T}^1$:
or by
standard continuum integration, or by doing a sum over points of integer
lattice,
each taken
with unit measure. For the torus with the perimeter equal $p$ (which is always
even)
there are exactly $(p/2)^2$ points from d.m.s. lying in ${\cal T}_{1,1}={\bf
T}^1$.
Thus the total volume per one copy of the initial moduli space is $(p/2)^2$
divided by
number of copies, i.e. $p^2/24$ in our case.

\subsection{General case of $\Mc$.}

The simplest example which has been considered in the previous subsection
permits us to
formulate the basic hypothesis about the structure of an arbitrary modular
space $\Mc$
in the Kontsevich parametrization. The main results we already have obtained
are:

{\bf 1.} From the comparison of two matrix models (Section~4) we got the
identity
(\ref{h13}) --- the indices for both continuum and d.m.s. coincide.

{\bf 2.} The calculation of all these indices  can be reduced to evaluating
the integrals
of the volume forms
over some finite coverings (simplicial complices) $\Tgn$ corresponding
to these modular spaces. In the discrete case values of these integrals are
just the
total numbers of integer points inside $\Tgn$ for some fixed values of
$p$--variables.
It means that all such points contribute to the sum with the same coefficient
--- the
volume of the unit cell. In its turn it means that {\it there should be no
points of
nonzero intrinsic curvature} inside $\Tgn$. In particular, all orbifold points
are not
singular points of metric in $\Tgn$.

{\bf 3.} All continuum modular spaces are closed compact manifolds without
boundaries
after compactification by Deligne and Mumford.

Thus we conclude that in the Kontsevich parametrization
for each modular space $\Mc$ we deal with the finite
covering $\Tgn$ which is totally conformally flat compact manifold without a
boundary.
The only possibility to ensure this property is to take $d$--dimensional
complex torus
${\bf T}^d$, $d=3g-3+n$. It should possesses a rich discrete group of
symmetries
$G_{g,n}$. Thus we conclude that in the Kontsevich parametrization $\Mc$ is
represented
as a factor over this symmetry group:
\be
\Mc={\bf T}^d/G_{g,n}.
\label{fact}
\ee
It is this relation which permits us to make further assumptions about
possibility to introduce quantum group
structure on these modular spaces.

\subsection{Proposal for Quantum Group Structure on $\Mcdisc$.}

All our previous considerations went along the line of the matrix model
theories. In
this subsection we want to leave the frames of these models and to speculate
about
possible underlying algebraical structures which can exist on d.m.s. We suppose
to
deal with this question in a special paper and restrict ourselves here by
making only
few notes.

First, let us have a closer look on the structure of d.m.s. for $\MOC$. Here we
again
reproduce the explicit form of the torus ${\bf T}^1/G_{1,1}$ for first few
values of
$p=2,4,6,8$ (Fig.8). We denote by small discs the points of
$\tpip[\Mdclose{1,1}]$.

\vspace{4.5cm}

\begin{picture}(0,0)(-20,0)
\multiput(20,20)(10,0){2}{\circle*{3}}
\multiput(25,28.33)(10,0){2}{\circle*{3}}
\multiput(20,20)(10,0){2}{\line(3,5){5}}
\multiput(20,20)(5,8.33){2}{\line(1,0){10}}
\put(30,20){\line(-3,5){5}}
\put(25,33){\makebox{$p=2$}}
\multiput(40,20)(10,0){3}{\circle*{3}}
\multiput(45,28.33)(10,0){3}{\circle*{3}}
\multiput(50,36.66)(10,0){3}{\circle*{3}}
\multiput(40,20)(10,0){3}{\line(3,5){10}}
\multiput(40,20)(5,8.33){3}{\line(1,0){20}}
\multiput(50,20)(15,8.33){2}{\line(-3,5){5}}
\put(60,20){\line(-3,5){10}}
\put(50,42){\makebox{$p=4$}}
\multiput(70,20)(10,0){4}{\circle*{3}}
\multiput(75,28.33)(10,0){4}{\circle*{3}}
\multiput(80,36.66)(10,0){4}{\circle*{3}}
\multiput(85,45)(10,0){4}{\circle*{3}}
\multiput(70,20)(10,0){4}{\line(3,5){15}}
\multiput(70,20)(5,8.33){4}{\line(1,0){30}}
\multiput(80,20)(30,16.66){2}{\line(-3,5){5}}
\multiput(90,20)(15,8.33){2}{\line(-3,5){10}}
\put(100,20){\line(-3,5){15}}
\put(85,50){\makebox{$p=6$}}
\multiput(110,20)(10,0){5}{\circle*{3}}
\multiput(115,28.33)(10,0){5}{\circle*{3}}
\multiput(120,36.66)(10,0){5}{\circle*{3}}
\multiput(125,45)(10,0){5}{\circle*{3}}
\multiput(130,53.33)(10,0){5}{\circle*{3}}
\multiput(110,20)(10,0){5}{\line(3,5){20}}
\multiput(110,20)(5,8.33){5}{\line(1,0){40}}
\multiput(120,20)(45,25){2}{\line(-3,5){5}}
\multiput(130,20)(30,16.66){2}{\line(-3,5){10}}
\multiput(140,20)(15,8.33){2}{\line(-3,5){15}}
\put(150,20){\line(-3,5){20}}
\put(130,60){\makebox{$p=8$}}
\multiput(175,35)(5,0){3}{\circle*{2}}
\end{picture}

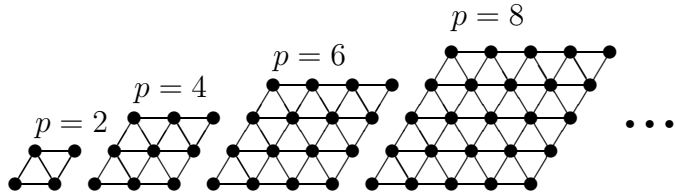
\centerline{Fig.8. Examples of orbits of $\tpip[\Mdclose{1,1}]$}

\vspace{8pt}

As $p\to\infty$ we reconstruct the space $\MOC={\bf T}^1/G_{1,1}$ for the
continuum
case. On the space $\MOC$ there exists Poisson structure due to Kontsevich
\cite{Kon91}. We expect that  there should be an analogue structure on each
orbit
$\tpip[\Mdclose{1,1}]$. Here the following question arises: which structures
except the
trivial one (the commutative group of shifts along the axis directions) would
exist for
such spaces? We know that there are quantum group structures for arbitrary
quantum tori
\cite{GolLeb} and we think that such the structures should exist for the case
under
consideration as well. It is reasonable to propose that the parameter $p$
labels
irreducible representations of some Quantum Group $Q_{1,1}$. Further, we claim
that
these representationa are excsptional, i.e. without highest weight vectors
(since our
torus space is the compact space without the boundary). There is no such
representations for the classical case, and in the quantum one they exist only
for
specific values of the quantum parameter $q=\e^{2\pi im/n}$. For such $q$
possible
dimensions of these irreducible exceptional representations belong to the
finite
series. For $SU_q(N)$ they are
$\{n^1,n^2,\dots,n^{N(N-1)/2}\}$ --- the highest power is the dimension of the
space of
the positive roots of the algebra. We see that it is compatible with our
consideration
if we assume $p=2n$ and as a proper candidat for $Q_{1,1}$ we have $SU_q(3)$.
Then it
is worth to note that the torus ${\bf T}^1$ we have is just the torus of the
Weyl
subgroup of $SU(3)$. The generators of $SU_q(3)$ should act on the space of
functions
$f[\tpip]$ and these representations should also respect the additional
symmetry transformation group $G_{1,1}$ which is a group of the sixth order in
our case.

We propose a hypothesis that it is a general situation for every d.m.s.
$\Mcdisc$.
{}From the formula (\ref{fact}) it is clear that at least such construction is
possible
because it is valid for any such torus and the only question is whether such a
representation exists which is invariant under the action of the symmetry group
$G_{g,n}$. I fit is a true, then
the d.m.s. are nothing but spaces on which the exceptional representations of
the Quantum Groups are living. We can also propose a hypothesis that for
the spaces with one puncure, $\Mdclose{g,1}$, these representations are of the
QG
$SU_q(2d+1)$, the torus ${\bf T}^d$ being in this case a Weyl subgroup torus of
this
Lie group.

Then the intersection indices may have an interpretation as traces of some
quantum
operators which do not depend on the representation chosen and even coincide
with the
one for a classical case. This question obviously deserves a big field for an
investigation.

\newsection{Conclusions}

In the conclusion we suppose to discuss briefly the perspectives
of the present action as well as the connection with another
approaches originated from the
matrix model technique.

First, as we see from our basic formula (\ref{h13}), for each moduli space
$\Mc$
there is a wide class of integrals which can be reduced to a sum over finite
sets of
points inside this space. It resembles a situation described in \cite{Niemi}
where the
hidden symmetry of functional inetgrals permits to reduce the procedure of
doing the
integral over the total space to taking the sum over singularity points. (We
owe to
A.Alexeyev by this note). So it is a reasonable task to search similar
structuires on
the moduli spaces themselves. Second, it is interesting to investigate the very
structure of moduli spaces using our main relation (\ref{dgen}). (We present an
example
of such calculation in the Appendix for the modular space $\Mcpar{2,1}$).

\newsection{Acknowledgements}

I am grateful to A.Alexeyev, D.Boulatov, G.Falqui, V.Fock, S.Frolov,
A.Gerasimov,
A.Lo\-s\-s\-ev, A.Marshakov,
A. Mironov, A.Morozov for numerous valuable discussions. I am
indebted to my collaborators J.Ambjorn, C.Kristjansen and Yu.Makeenko. I thank
L.D.Fa\-d\-deev, A.Kirillov and A.Venkov for their interest to my work. I would
like to
thank Prof. P. Di Vecchia and
NORDITA for hospitality during my visit to Copenhagen where
part of this work was done. I am grateful to Prof. P.K.Mitter and LPTHE for
permanent
support of my action and for excelent conditions in Paris where this work has
been
completed.

\newsection{Appendix. The explicit solution to $\Mcpar{2,1}$}

Here we shall find by the direct calculation the form of the formula
(\ref{dgen}) for
the case of genus two moduli space with one puncture. In the paper \cite{ACKM}
the
explicit form of genus two partition function in terms of momenta was found:
\bea
F_2&=&-\frac{119}{7680J_1^2d^4}-\frac{119}{7680M_1^2d^4}+\frac{181J_2}{480J_1^3d^3}
-\frac{181M_2}{480M_1^3d^3}\nonumber\\
&{}&+\frac{3J_2}{64J_1^2M_1d^3}-\frac{3M_2}{64J_1M_1^2d^3}-\frac{11J_2^2}{40J_1^4d^2}
-\frac{11M_2^2}{40M_1^4d^2}\nonumber\\
&{}&+\frac{21J_2^3}{160J_1^5d}-\frac{21M_2^3}{160M_1^5d}-\frac{29J_2J_3}{128J_1^4d}
+\frac{29M_2M_3}{128M_1^4d}\nonumber\\
&{}&+\frac{35J_4}{384J_1^3d}-\frac{35M_4}{384M_1^3d}.
\label{aa1}
\eea
In order to investigate the modular space $\Mcpar{2,1}$ it is enough to use the
expansions (\ref{h12}), because we must keep only terms of the first order in
traces.
Then the only thing we need more is to express the quantities
\bea
p_k&=&\frac{\e^{\l(k+1)}}{(\eL -1)^{2k+1}(\eL +1)},\nonumber\\
q_k&=&\frac{\e^{\l(k+1)}}{(\eL -1)(\eL +1)^{2k+1}}
\eea
via the derivatives $L_a$ and $R_a$ (\ref{h7}) using (\ref{h9}) and
(\ref{h11}).
We omit all lenghty calculations and present here only the final answer. After
replacing $\aa\to -\aa/2$ we remain with
\bea
&{}&\phantom{XXXXX}
w_2(\l)=\frac{1}{2^d}\Biggl\{\frac{L_4}{4!}\cdot\frac{1}{1152}-\frac{L_3}{3!}\cdot
\frac{7}{1152}\Biggr.\nonumber\\
&{}&+\Biggl.\frac{L_2}{2!}\cdot\frac{1}{1152}\cdot\frac{1781}{30}-
L_1\cdot\frac{1}{1152}\cdot
\frac{32581}{480}+L_0\cdot\frac{119}{192\cdot 40}+(L_a\to R_a)
 \Biggr\}.
\label{aa2}
\eea
Here $\frac{1}{1152}=\<\tau_4\>_2^{}$ and we see that the sum over reductions
appear to
give highly nontrivial coefficients. One should take into account that only for
the
first reduction degree there is unique diagramm which correspond to, for all
higher
reductions there are various kinds of reductions, hence, the coefficients
standing with
the terms in the second line of (\ref{aa2}) are themselves to be presented as
sums of
symmetrical coefficients, each corresponding to some selected type of the
reduction.

\end{document}